\documentclass[aps,prd,nofootinbib,letterpaper]{revtex4}
\usepackage{amsmath,amssymb}
\usepackage[usenames,dvipsnames]{color}
\usepackage{graphicx}
\usepackage{float}
\usepackage{geometry} 
\usepackage{epstopdf}
\usepackage{tensor,physics}
\usepackage[bookmarks]{hyperref}
\usepackage{natbib}
\begin{document}

\title{Fixing extensions to general relativity in the nonlinear regime}
\author{Juan Cayuso}
\email{jcayuso@perimeterinstitute.ca}
\author{N\'estor Ortiz}
\email{nortiz@perimeterinstitute.ca}
\author{Luis Lehner}
\email{llenher@perimeterinstitute.ca}
\affiliation{Perimeter Institute for Theoretical Physics, 31 Caroline Street, Waterloo, Ontario N2L 2Y5, Canada}
\date{\today}

\begin{abstract}
The question of what gravitational theory could supersede General Relativity has been central in theoretical physics for decades. Many disparate alternatives have been proposed motivated by cosmology, quantum gravity and phenomenological angles, and have been subjected to tests derived from cosmological, solar system and pulsar observations typically restricted to linearized regimes.  Gravitational waves from compact binaries provide new opportunities to probe these theories in the strongly gravitating/highly dynamical regimes. To this end however, a reliable understanding of the dynamics in such a regime is required. Unfortunately, most of these theories fail to define well posed initial value problems, which prevents at face value from meeting such challenge. In this work, we introduce a consistent program able to remedy this situation. This program is inspired in the approach to ``fixing'' viscous relativistic hydrodynamics introduced by Israel and Stewart in the late 70's. We illustrate how to implement this approach to control undesirable effects of higher order derivatives in gravity theories and argue how the modified system still captures the true dynamics of the putative underlying theories in 3+1 dimensions. We sketch the implementation of this idea in a couple of effective theories of gravity, one in the context of Noncommutative Geometry, and one in the context of Chern-Simons modified General Relativity.
\end{abstract}

\maketitle
\section{Introduction}
Understanding the nature of gravity, and what should supersede General Relativity (GR) to describe it, is one of the most important questions in modern theoretical research. Efforts stemming from cosmology, quantum gravity and particle physics have provided a plethora of interesting suggestions for possible extensions. Many of such ideas have been confronted with exquisite cosmological, solar system, and pulsar tests, which have helped constrain and further shape or discard them (e.g.~\cite{Will:2014bqa,Freire:2012mg,Baker:2012zs,Yunes_Siemens13,Berti_etal15}). The recent start of gravitational wave astronomy is providing the opportunity to explore such theories in the previously inaccessible strongly gravitating/highly dynamical regime. Indeed, the detection of gravitational waves emitted by binary black hole systems has already provided the first tests of GR~\cite{Abbott_etal16,Yunes_etal16}, which will be further refined as more events become available, e.g. \cite{Cornish:2011ys,Meidam:2014jpa,Sampson:2014qqa,Yang:2017zxs}. Incipient efforts are attempting to map out predictions from extensions/alternatives to GR that can be confronted with observations. For this purpose, specific predictions from such theories for compact binary mergers are required but, unfortunately, this has so far only been possible in a (very) small number of  theories~\cite{Berti_etal15,Yunes_Siemens13}. One main reason for this is the fact that most extensions to GR lead to initial value problems whose well-posedness is, at best, suspect. As first stated by Hadamard, well-posedness is a requirement for any physical theory in a way that its field equations possess a unique solution which depends continuously on the initial data~\cite{jH02,Gustafsson1995}. In particular, a large class of extensions to GR contemplate field equations which involve higher order derivatives, stemming from effective Lagrangians involving higher order curvature contributions, which, in turn, lead to ghost degrees of freedom, acausal propagation of perturbations, and generically runaway energy cascades to the ultraviolet (as an example, Ref.~\cite{Papallo:2017qvl} discusses how generic ill-posedness in Lovelock and Horndeski theories is). At the linearized level, the aforementioned issues are typically ``controlled'' by suitably imposed cutoffs, explicit removal of ghosts modes, etc. At nonlinear regimes however, these strategies can not be readily applied. Moreover, since such regimes typically require complex numerical simulations---which seed all possible frequencies with truncation errors---a well posed problem is a necessary condition for obtaining reliable predictions~\cite{Gustafsson1995}. Faced with this situation, some type of pragmatical approach should be considered to explore nonlinear regimes and some possible first steps have recently been  introduced ~\cite{Endlich:2017tqa,Okounkova:2017yby}. Whichever the strategy is, it is important to stress that it should remain as faithful as possible to the motivating principles employed to construct such GR extensions. Otherwise, solutions obtained with a given strategy need not be representatives of the phenomena in the extended gravitational theory under consideration.

\begin{figure}[h!]
\begin{center}
\includegraphics[width=9cm]{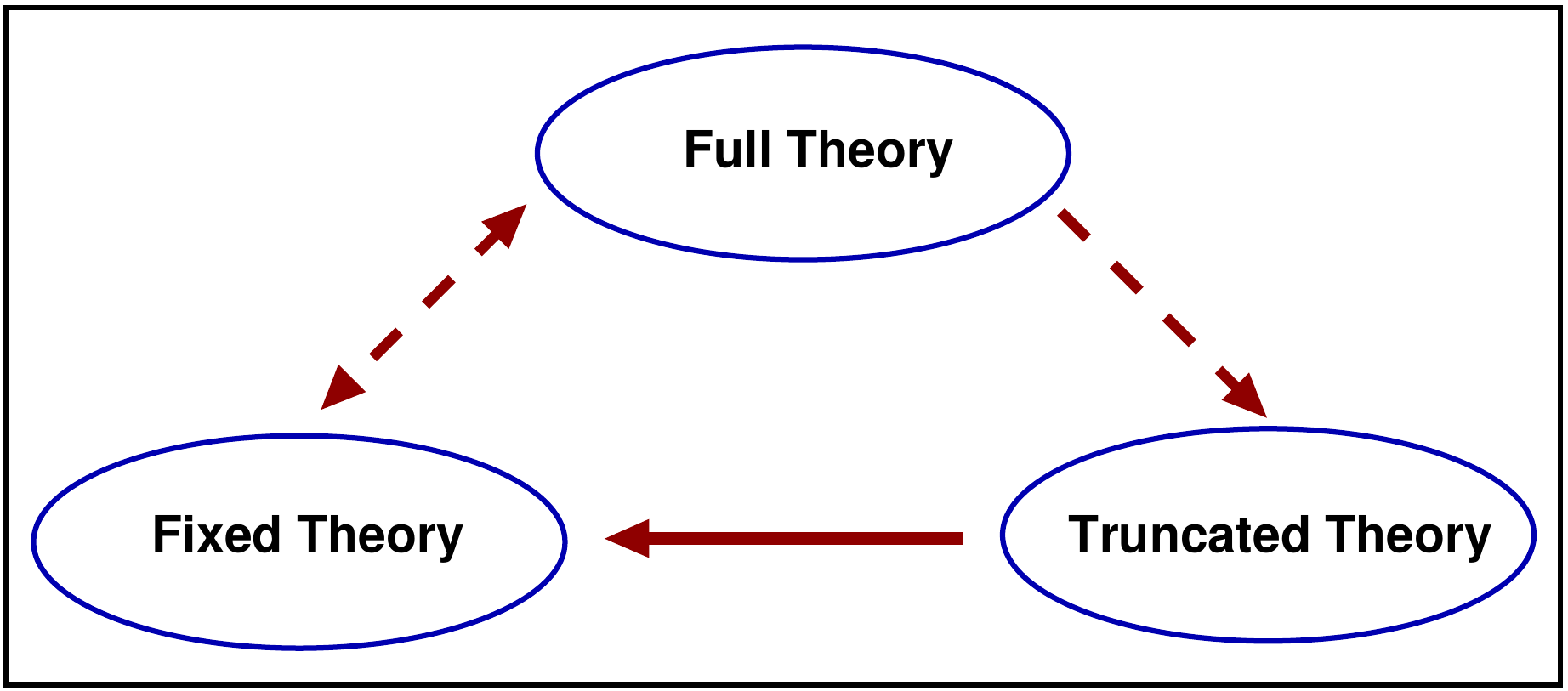}
\end{center}
\caption{\label{Fig:Diagram} (Color online). Schematics of the three possible theories. First, a {\em Full Theory} is envisioned which is free of pathologies and, in particular, problems studied in it are well posed. Such theory is in general unknown and a {\em Truncated Theory} is written instead. Ideally, phenomenology/observations drawn from the Truncated Theory are expected to guide the construction of the Full Theory. Since the Truncated Theory typically has pathologies, a {\em Fixed Theory} is to be constructed. When done consistently, the Fixed Theory informs about the Truncated Theory and, in turn, the Full Theory. }
\end{figure}

In the current work, our aim is not to survey the whole gamut of extensions to GR but instead focus on broadly applicable statements, and present a program to remedy shortcomings in the exploration of such theories within gravitational wave astronomy, in particular allowing for capturing the full dynamics of the system within the regime they were written for in 3+1 dimensions. This program can then be appropriately incorporated in the extension of choice. As a result, a {\em fixed theory} is introduced as a suitable modification of the {\em truncated theory}. In the fixed theory, as opposed to the truncated one, the equations of motion yield a well posed problem and thus allow one to study, in particular, their nonlinear behavior. This is schematically represented by the solid arrow in Fig.~\ref{Fig:Diagram}. As we shall discuss, our strategy to construct a fixed theory relies on two foundations. The first one guides how to {\em modify the equations of mnonlinearotion} so that they lend themselves to defining well posed problems for the regime of interest. These modifications enforce a consistent gradient expansion, and thus truncation of the underlying equations is maintained, and consequently  energy cascade to the ultraviolet is controlled. The methodology is inspired in the so-called Israel-Stewart formulation of relativistic viscous hydrodynamics which  addressed similar problems. The second one argues {\em one indeed obtains reliable answers in  3+1 dimensions} after suitably modifying the dynamics as described above. Such argument relies on indications that GR---and viable extensions---in the long wavelength regime of perturbations to black hole spacetimes, favours an indirect energy cascade. At the theoretical level, such indications are motivated by the fluid/gravity correspondence in Anti-de Sitter (AdS) spacetime~\cite{Sayantani_etal08,Sayantani_etal08b,VanRaamsdonk08}, the study of perturbations of highly spinning black holes in asymptotically flat spacetimes~\cite{Yang_etal15}, arguments  from nonlinear couplings in General Relativity~\cite{Yang:2015jja,Galtier:2017mve}, and, most importantly, by the lack of significant short-scale features in observed gravitational waves of merging black holes. Supported by these indications, reliability on calculations obtained from ``fixed" theories, would ultimately allow confrontation of solutions versus actual phenomena observed in the nonlinear regime. This is represented by the left, dotted arrow closing the cycle of Fig.~\ref{Fig:Diagram}.
This work is organized as follows. In Sec.~\ref{Sec:IS} we succinctly review the Israel-Stewart formulation of relativistic viscous hydrodynamics, and we argue in support of its application to gravity theories. In Sec.~\ref{Sec:Warmup} we provide an example of the implementation of our approach in a particular toy model. In Sec.~\ref{Sec:Noncommutative} we apply the idea to the case of linearized gravity in the context of Noncommutative Geometry. In Sec.~\ref{Sec:CS}, the same idea is implemented in a couple more models: The first one is inspired on the linearized field equations of dynamical Chern-Simons gravity, whereas the second one is inspired in the Cauchy formulation of the same field equations. We conclude in Sec.~\ref{Sec:Conclusions}. Through this manuscript, early latin symbols $a, b, c, ...$ correspond to spacetime indices, whereas middle latin ones $i, j, k, ...$ correspond to spatial indices. We work with 
geometrized units ($G=c=1$).

\section{Brief review on Israel-Stewart's formulation of Relativistic Viscous Hydrodynamics}\label{Sec:IS}
To fix ideas, we step back from analyzing the case of gravity and first review the formulation of relativistic viscous hydrodynamics, which, as we shall see, bears a close relation to our goal. For concreteness and without loss of generality, we will concentrate on neutral, conformal fluids described by a trace-free stress-energy tensor $T_{ab}$ on a $d$-dimensional flat background. We adopt the modern approach to relativistic hydrodynamics as an expansion in gradients of the fluid fields~\cite{Baier_etal08,Romatschke10}, and keep terms to second order in these gradients. This is then a {\em truncated theory} which, as described next, has pathologies when dissipation is considered.
In concrete terms, for a conformal fluid of total energy density $\rho$ and pressure $p$ measured by observers comoving with $d$-velocity $u^a$, we consider the expansion
\begin{eqnarray}
T_{ab} &=& (\rho+p) u_a u_b + p \eta_{ab} \nonumber \\
&-& \eta \sigma_{ab} + \eta \tau_{\Pi} \left[ {}_{<}D\sigma_{ab>} + \frac{1}{d-1} \sigma_{ab} \nabla_c u^c \right] \nonumber \\
&+& \lambda_1 \sigma_{c<a} \sigma_{~b>}^c + \lambda_2 \sigma_{c<a} \Omega_{~b>}^c
+ \lambda_3 \Omega_{c<a} \Omega_{~b>}^c, \label{stressfluid}
\end{eqnarray}
where $D := u^a\nabla_a$ is the derivative operator in the direction of $u^a$, and $\Omega^{ab} := h^{ac} h^{bd} \nabla_{[c} u_{d]}$ and $\sigma_{ab}/2 := {}_{<} \nabla_{a} u_{b>}$ are the fluid's vorticity and shear tensors, respectively, defined by means of the projector to the space orthogonal to $u^a$, $h_{ab} := \eta_{ab} + u_a u_b$, and the symmetric traceless part of such projection, which for a second rank tensor $A_{ab}$ is given by
\begin{equation}
{}_<A_{ab>} := h_{ac} h_{bd} A^{(cd)} - \frac{1}{d-1} h_{ab} h_{cd} A^{cd}.
\end{equation}
In expansion~(\ref{stressfluid}), $\eta$ represents the fluid's shear viscosity, whereas $(\tau_\Pi,\lambda_1, \lambda_2, \lambda_3)$ denote nonlinear transport coefficients.
The fluid's equations of motion are obtained from the conservation laws $\nabla_a T^{ab}=0$ coupled to the constraint $u^a u_a = -1$ and the equation of state $p =\rho/(d-1)$ which is imposed by conformal invariance.
As is well known, restricting to the perfect fluid case---first line of Eq.~(\ref{stressfluid})---yields a strongly hyperbolic system of equations which, supplemented with appropriate initial and boundary conditions, yields a well posed Cauchy problem. In turn, the inclusion of viscous terms breaks this property, hyperbolicity is lost and features like acausal propagation of perturbations due to unbounded phase velocities appear, as well as runaway energy transfer~\cite{Buchel:2009tt,Romatschke10}. Such unphysical behavior is understood as one regards relativistic hydrodynamics as an effective theory in the long wavelength regime. Thus, as energy runs towards shorter wavelengths due to nonlinearities, one abandons the regime of applicability of the gradient expansion, and the evolution system for the hydrodynamic variables can no longer be trusted. Although at the linear level this could be dealt with by imposing a physical ultraviolet cutoff, such strategy is not well justified in the nonlinear regime. Unfortunately, in general cases, the full theory is unknown, leaving us with no theory to explore all the regimes of interest.

One option to remedy this situation was introduced by Israel and Stewart~\cite{Israel76,Israel_Stewart76,Israel_Stewart79}. They derived alternative equations of motion for relativistic viscous hydrodynamics starting from {\it extended irreversible thermodynamics} considerations~\cite{Jou_etal98}. The resulting set of equations, which govern the {\em fixed theory}, is constructed so as to {\em ensure} staying within the regime of applicability of the {\em truncated theory}. The underlying assumption is that a gradient expansion is applicable in the physical problem of interest. If this condition is not satisfied, then one is outside the regime of applicability of the truncated theory. For a complete, didactic derivation of the Israel-Stewart theory, we refer to~\cite{Romatschke10}, for instance. For brevity, here we describe a rather pragmatic approach, which can be succinctly described as follows. First, promote everything but the perfect fluid part of $T_{ab}$ as a new tensor $\Pi_{ab}$---namely the {\it viscous part} of the stress-energy tensor. Next, notice that, to leading order, $\Pi_{ab} = - \eta \sigma_{ab}$. Then, one can write
\begin{eqnarray}
\Pi_{ab} &=& - \eta \sigma_{ab} \nonumber \\
& & - \tau_{\Pi} \left[ {}_{<}D\Pi_{ab>} + \frac{1}{d-1} \Pi_{ab} \nabla_c u^c \right]
+ \frac{\lambda_1}{\eta^2} \Pi_{c<a} \Pi_{~b>}^c - \frac{\lambda_2}{\eta} \Pi_{c<a} \Omega_{~b>}^c
+ \lambda_3 \Omega_{c<a} \Omega_{~b>}^c. \label{higher_stressfluid}
\end{eqnarray}
One can then turn this equation into an evolution equation for $\Pi_{ab}$ of the
form
\begin{equation}\label{Eq:IS_modification}
\tau_{\Pi} u^c \nabla_c \Pi_{ab} = -\Pi_{ab} - \eta \sigma_{ab} + \mbox{lower~order~terms},
\end{equation}
which ensures that in a timescale given by $\tau_{\Pi}$, $\Pi_{ab}$ approaches $\eta \sigma_{ab}$, thus higher derivatives are sub-leading. The resulting system of equations is again strongly hyperbolic, and well posed problems can be defined. Mathematically, the problem is now well defined; but what about the physics? That is, has this modification altered the physics obtainable with these equations in such a way
that their results bear little (or limited) relation to the original system---as long as such system can reliably treat problems of interest? We recall that the {\em fixed theory} has been written to enforce a gradient expansion condition. As a result, energy flow to higher frequency modes is reduced. Is this consistent with the desired physics? Interestingly, the answer to this question depends on the dimensionality of the problem. Indeed, in two spatial dimensions---where energy preferentially flows to lower frequency modes---the {\em fixed theory} does not affect the outcome in a significant way when applied to initial conditions within the regime of applicability of the gradient expansion. This behavior is due to the existence of a (quasi) conserved quantity dubbed as {\em enstrophy}~\cite{Kraichnan67}. The fact that the rate of change of enstrophy is low constrains the amount of energy that can flow to high frequencies~\cite{Carrasco_etal12,Westernacher-Schneider_Lehner15}. Consequently, the Israel-Stewart formulation applied to 2+1 viscous relativistic hydrodynamics yields solutions which should approximate well the underlying (putative) theory\footnote{In practical terms, solutions in 2+1 dimensions do not depend sensitively on the value of $\tau_{\Pi}$~\cite{Green_etal14}.}. Its main effect is simply to control high frequency modes which, especially in computational implementations, can be easily triggered~\cite{Hiscock_and_Lindblom83}. As mentioned then, applying the Israel-Stewart formulation of viscous relativistic hydrodynamics does not only lead to well posed problems but, in 2+1 dimensions, provides reliable answers with respect to the putative full theory~\cite{doi:10.1063/1.530958}. Likewise, when addressing extensions to  General Relativity derived from an effective theory approach, the assumption is that the problem of interest is in the regime of applicability of the truncated theory. Next, we will discuss scenarios when staying in such regime is likely.

\subsection{Possible implications on gravity}\label{SubSec:Implications}
The above discussion highlighted two important aspects that we will exploit in our program. First, what the Israel-Stewart approach to ``fixing'' viscous relativistic hydrodynamics entails mathematically. Second, that such fix should provide faithful solutions to the case of 2+1 dimensions for the hydrodynamical problem. Let us comment on the second property and its connection for our purposes in gravitational theories.

\begin{itemize}
\item In the context of long-wavelength perturbations of asymptotically AdS black holes, a fluid/gravity correspondence has been derived~\cite{Sayantani_etal08,Sayantani_etal08b,VanRaamsdonk08}, which uniquely relates $d$-viscous relativistic hydrodynamics with $d+1$ gravity. This correspondence shows the intrinsic relation of gravitational and hydrodynamical solutions in perturbations with long wavelengths. In particular, 2+1 hydrodynamics phenomena has a counterpart in 3+1 gravity (e.g.~\cite{VanRaamsdonk08,Eling_etal10,Carrasco_etal12})\footnote{When applied to 3+1 hydrodynamics, the Israel-Stewart formulation produces solutions that do depend sensitively on the additional parameters introduced (e.g. $\tau_{\Pi}$), thus it requires a thorough analysis to understand the extent and regime in which solutions can indeed be trusted. The same caveat applies if our program were to be implemented in higher dimensional theories of gravity (unless such dimensions were to be small or somehow screened).}. Furthermore, arguments for the existence of a gravitational analogue to enstrophy have been provided as well as direct simulations of the inverse energy cascade in both hydrodynamics and gravitational problems~\cite{Eling_etal10,Carrasco_etal12,Adams_etal14,Green_etal14,Westernacher-Schneider_Lehner15,Westernacher2}.
\item In the context of long-wavelength perturbations of highly spinning black holes in
asymptotically flat spacetimes, recent work has revealed a close connection between {\em gravitational turbulence} and the analog phenomena in hydrodynamics~\cite{Yang_etal15}.
As well, tantalizingly, the Bianchi identities at future null infinity, when expressed in terms of the vorticity field~\cite{Green_etal14} reduces
to an equation analog to the one governing the vorticity in hydrodynamics, from which the slow rate of change of 
enstrophy is understood.
\item The recent detection of gravitational waves of merging black holes has shown waveforms that do
not indicate a significant transfer of energy to shorter wavelengths takes place,
even when the spacetime around a given black hole is perturbed in a significant way through the merger with 
another~\cite{Abbott:2016blz,Abbott:2016nmj,Abbott:2017}.
\end{itemize}

The above considerations, coupled with the celebrated result of the stability of Minkowski spacetime for initial data in a suitable Sobolev norm~\cite{Christodoulou_Klainerman93,Lindblad_Rodnianski10}---which limits the amount of energy that higher frequency modes acquire---suggest that, for long-wavelength perturbations, 3+1 perturbations do not have a significant direct energy cascade. Consequently, it is natural to expect that higher derivatives present in the equations of motion of extensions to GR, in the right regimes, could be well behaved. At this point we should stress that this is not always the case.
It depends on the scenario under consideration.
For example, strong energy cascades to shorter wavelengths occurring in the context of the singularity theorems~\cite{1965PhRvL..14...57P,Hawking:1973uf}, would make our approach not applicable. Another relevant example would be black hole formation. In the case of gravitational collapse of a compact star to a black hole, for instance, our approach might be applicable as long as higher order gradients do stay reasonably small---which is indicated by simulations in GR. However, in critical collapse~\cite{mC93}, it will not be applicable.

\section{General idea and toy model example}\label{Sec:Warmup}
In order to illustrate how an approach inspired in the Israel-Stewart strategy can help to alleviate the blowing-up of solutions and unbounded phase velocities of high frequency modes, in this section we introduce a toy model capturing these issues. Let us consider the following non-linear partial differential equation for a complex scalar field $\psi$ defined in $1+1$ Minkowski spacetime in Cartesian coordinates $(t,x)$, 
\begin{equation}\label{Eq:NonLinear}
\Box \psi = \psi\partial_{x}^{n} \psi,
\end{equation}
where $\Box := -\eta^{\mu\nu}\nabla_\mu\nabla_\nu$ is the d'Alembert operator, and $\partial_x^n$ indicates $n$-th order differentiation with respect to the spatial coordinate $x$.
In order to explore the possible non-linear phenomenology of Eq.~(\ref{Eq:NonLinear}), we follow the standard approach of linearizing it with respect to a general
solution and study the behavior of general perturbations. As we will show in Sec.~\ref{Sec:Noncommutative}~and~Sec.~\ref{Sec:CS}, linearized solutions of effective gravity theories share features with linearized solutions of the toy model Eq.~(\ref{Eq:NonLinear}). We first notice that $\psi = \mbox{const.}$ is a trivial solution of the nonlinear Eq.~(\ref{Eq:NonLinear}). Let us take $\lambda\in\mathbb{R}^+$ and consider the linearization
\begin{equation}\label{Eq:Warmup_linealization}
\psi=\lambda+\phi,
\end{equation}
where the complex function $\phi$ represents a small deviation from the constant $\lambda$. Then, from Eq.~(\ref{Eq:NonLinear}) one obtains that, to leading order, $\phi$ satisfies
\begin{equation}\label{Eq:Warmup}
\Box \phi = \lambda \partial_{x}^{n} \phi.
\end{equation}
In order to analyze the solutions' behavior of the linear Eq.~(\ref{Eq:Warmup}), we consider the ansatz
\begin{equation}\label{Eq:Ansatz}
\phi(t,x) = A\exp(s t + i k x),
\end{equation}
with $A, k \in \mathbb{R}$, $s \in \mathbb{C}$, for which Eq.~(\ref{Eq:Warmup}) implies the ``dispersion relation" $s(k)~=~\pm\sqrt{\lambda(ik)^n-k^2}$. It is clear that $\lambda$ being positive allows for $s(k)$ to generically develop a positive, unbounded real part, which then gives rise to solutions with unbounded growth rates---which stands in the way of well-posedness. In the following Subsections we discuss the representative cases of $n=4$ and $n=3$.
\subsection{Case $n=4$}
In the case of $n=4$, the dispersion relation associated to Eq.~(\ref{Eq:Warmup}) under the ansatz in~Eq.(\ref{Eq:Ansatz}), is given by $s(k)~=~\pm ik\sqrt{1-\lambda k^2}$, revealing the existence of blowing-up modes for sufficiently large values of $k$---high frequencies. Moreover, for arbitrarily large values of $k$, the phase speed $||s(k)/k||$ approaches $\sqrt{\lambda}k$, which exceeds the speed of light for $k > 1/\sqrt{\lambda}$. This implies acausal mode propagation.

Let us now introduce the independent dynamical variable $\Pi := \partial_{x}^{2} \phi$, together with the {\it ad-hoc} evolution equation $\sigma\Box \Pi  =\partial_{x}^{2} \phi-\tau \partial_t \Pi - \Pi $, where $\sigma,\tau \in \mathbb{R}^+$. Our guiding principle in the construction of this equation for $\Pi$, is that it should keep high order gradients small by enforcing $\Pi$ to approach $\partial_{x}^{2} \phi$ in a timescale $\tau$. In general, the specific form of this equation would not matter as long as energy does not significantly cascade to shorter wavelengths. If energy flows strongly to higher frequencies, then the specific choice of evolution equation, which will have different inherent timescales will likely have a significant impact on the solution. After the introduction of $\Pi$ and its evolution equation, the original Eq.~(\ref{Eq:Warmup}) is replaced by the system
\begin{subequations}\label{Eq:Warmup_system}
\begin{eqnarray}
\Box \phi &=& \lambda \partial_{x}^{2} \Pi, \\
\sigma\Box \Pi  &=&\partial_{x}^{2} \phi-\tau \partial_t \Pi - \Pi.
\end{eqnarray}
\end{subequations}
As we shall show, provided that $\sigma$ is sufficiently large, namely $\sigma \geq \lambda$, solutions of Eqs.~(\ref{Eq:Warmup_system}) of the form $\phi(t,x) = A\exp(s t + i k x)$, $\Pi(t,x) = A\beta(s,k)\exp(s t + i k x)$, possess no blow-up modes, and their phase velocity is bounded for all $k$. In practice, nonexistence of blow-up modes is equivalent to nonpositivity of the real part of the dispersion relation solutions $s(k)$. In order to show this, we first notice that, with the given ansatz, Eqs.~(\ref{Eq:Warmup_system}) imply the algebraic system
\begin{subequations}
\begin{eqnarray}
\Big{(}s(k)^2 + k^2\Big{)}&=&-\lambda\beta(s,k)k^2, \\
\sigma\beta(s,k)\Big{(}s(k)^2 + k^2\Big{)}&=&-k^2-\tau\beta(s,k)s(k)-\beta(s,k),
\end{eqnarray}
\end{subequations}
from which, eliminating $\beta$, we obtain the quartic equation for $s(k)$,
\begin{eqnarray}\label{Eq:Quartic}
\Big{(}s(k)^2 + k^2\Big{)}\bigg{[}\sigma\Big{(}s(k)^2 + k^2\Big{)}+{\tau}s(k)+1\bigg{]}={\lambda}k^4.
\end{eqnarray}
We shall show that the solutions of Eq.(\ref{Eq:Quartic}) have nonpositive real part. For clarity we split the proof into two parts. In the first one, we show that $\Re[s(k)] \ne 0$ for $k \ne 0$. In the second one, we show that $\Re[s(0)] \leq 0$ and that it is a local maximum of $\Re[s(k)]$. Combining results, and by virtue of the continuity of $s(k)$, we will conclude that $\Re[s(k)] \leq 0$ for all $k \in \mathbb{R}$.\\

$\bullet$ For the first part of the proof, we observe that $s(k)=0$ is not a solution for $k\neq0$, the reason being clear if we rearrange Eq.~(\ref{Eq:Quartic}) as
\begin{equation}\label{Eq:simple11}
s(k)^2+{\tau}s(k)\Big{(}s(k)^2+k^2\Big{)}+{\sigma}s(k)^4+2{\sigma}s(k)^2k^2 = k^2 \left[ (\lambda-\sigma)k^2-1 \right],
\end{equation}
where $\sigma\geq\lambda$ guarantees nonvanishing of the right-hand side. Thus $s(k)\neq0$ for $k\ne0$.
Now we proceed by contradiction. Let $k\ne0$, $s(k) = u(k) + iv(k)$, and suppose $u(k)=0$, then $v(k)\ne0$, otherwise $s(k)=0$. 
Substituting $s(k)=iv(k)$ in Eq.~(\ref{Eq:simple11}) results into
\begin{subequations}
\begin{eqnarray}
v(k)^2\left[ {\sigma}v(k)^2 - 2{\sigma}k^2 -1 \right] &=& k^2 \left[ (\lambda-\sigma)k^2-1 \right], \label{Eq:proof1_a} \\
i{\tau}v(k)\Big{[}k^2 - v(k)^2 \Big{]} &=& 0.\label{Eq:proof1_b}
\end{eqnarray}
\end{subequations}
Equation~(\ref{Eq:proof1_b}) reveals $v(k)=\pm k$, which in Eq.~(\ref{Eq:proof1_a}) implies $\lambda = 0$, contradicting the hypothesis $\lambda \in \mathbb{R}^+$. Therefore $u(k) = \Re[s(k)] \ne 0$ for $k \ne 0$.\\

$\bullet$ For the second part of the proof, consider $k=0$. In this case, Eq.~(\ref{Eq:simple11}) reads
\begin{equation}
s(0)^2\Big{[}{\sigma}s(0)^2+{\tau}s(0)+1\Big{]}=0, \nonumber 
\end{equation}
which has the trivial solution $s(0)=0$, as well as
\begin{equation}
s_{\pm}(0)=-\frac{\tau}{2\sigma}\pm\sqrt{\Big{(}\frac{\tau}{2\sigma}\Big{)}^2-\frac{1}{\sigma}}.
\end{equation}
Thus $\Re[s(0)]\leq0$ regardless of the relation between $\tau$ and $\sigma$. Furthermore, $\Re[s(k)]$ reaches a local maximum at $k=0$, as can be inferred by differentiating Eq.~(\ref{Eq:simple11}) with respect to $k$ and evaluating in $k=0$, giving
\begin{equation}
s^{(1)}(0) = 0, \qquad
s^{(2)}(0) = i, \qquad
s^{(3)}(0) = -i3\lambda, \qquad
s^{(4)}(0) = -12\lambda\tau,
\end{equation}
where $s^{(i)}$ denotes the $i$-th derivative of $s$ with respect to $k$. Therefore, $\Re[s(k)] \leq 0$ for all $k \in \mathbb{R}$, which completes the proof.

\begin{figure}[h!]
\centering
\begin{minipage}{.5\textwidth}
  \centering
  \includegraphics[width=0.95\linewidth]{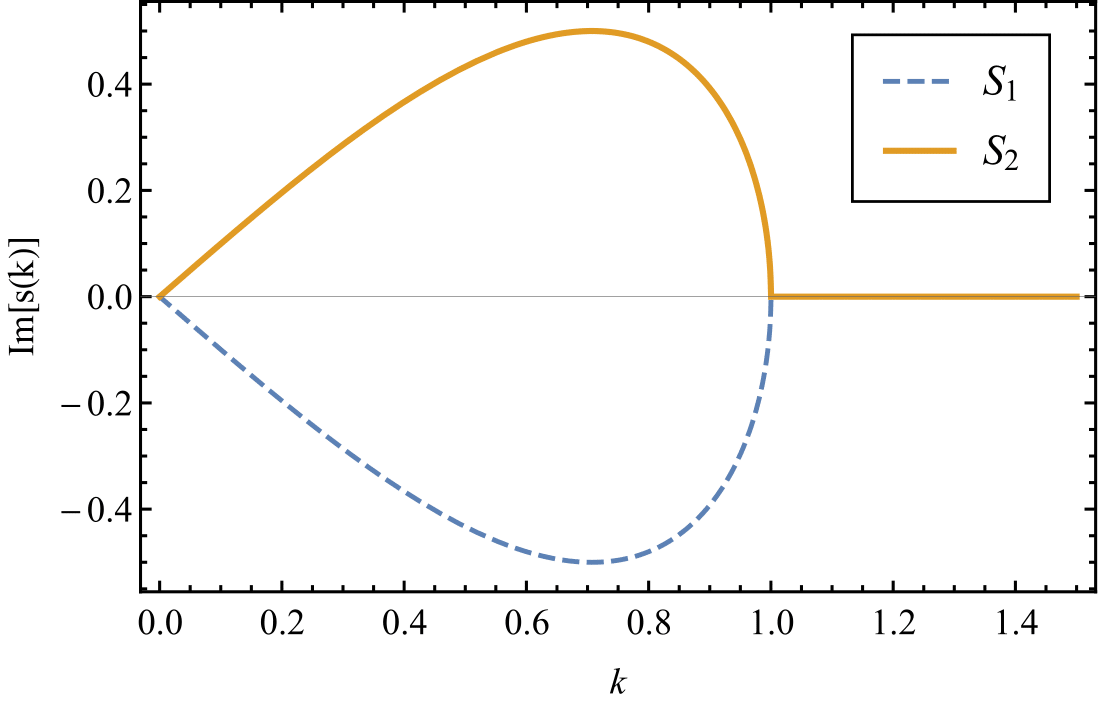}
\end{minipage}%
\begin{minipage}{.5\textwidth}
  \centering
  \includegraphics[width=0.95\linewidth]{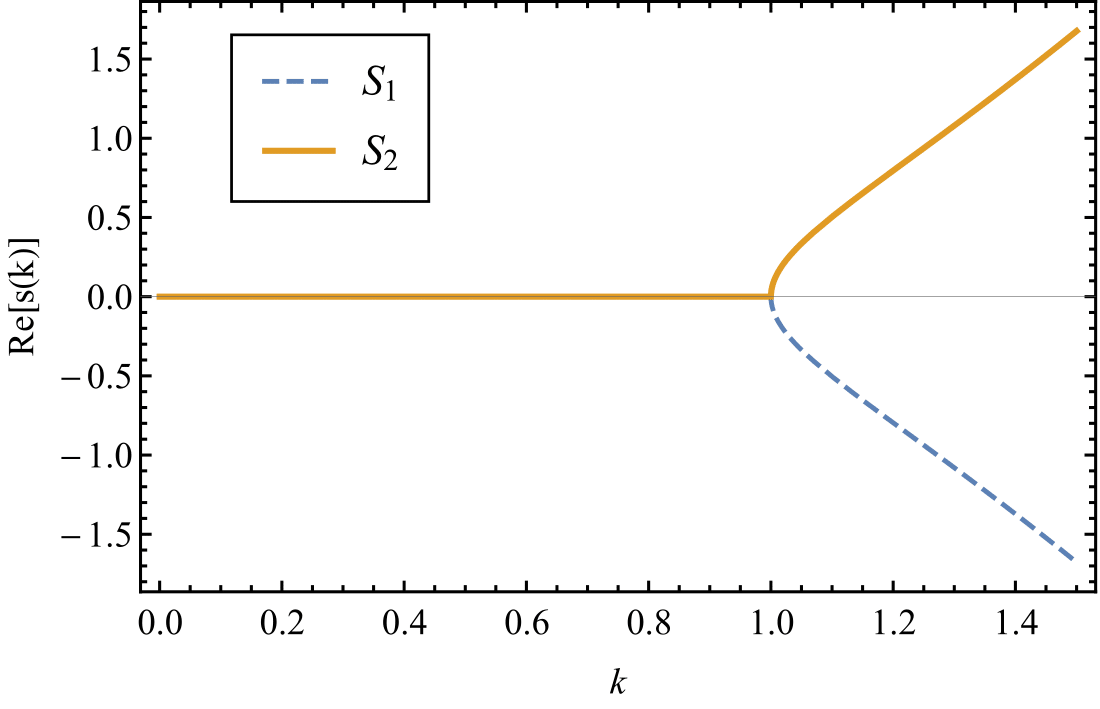}
\end{minipage}
\caption{\label{Fig:Warmup_original} (Color online) Real and imaginary parts of the dispersion relation solutions $s(k)$ for the original, linearized model Eq.~(\ref{Eq:Warmup}) with $n=4$ and $\lambda=1$.}
\end{figure}

In summary, the system of equations~(\ref{Eq:Warmup_system}) constitutes a modification of the original Eq.~(\ref{Eq:Warmup}) inspired on the Israel-Stewart approach to relativistic hydrodynamics. This modification is such that the dispersion relation $s(k)$ associated to the system~(\ref{Eq:Warmup_system}) has nonpositive real part, preventing the existence of blowing-up modes originally present in Eq.~(\ref{Eq:Warmup}).
Nevertheless, it is important to verify whether or not the modified system~(\ref{Eq:Warmup_system}) preserves the dynamics of the original Eq.~(\ref{Eq:Warmup}) in the low frequencies regime, as such is the regime that might have inspired
the model (or its truncation). Despite the fact that we have eliminated the troublesome blowing-up modes, we could have ``spoiled" the physics in the process. In order to investigate this possibility, we have numerically solved the polynomial equations for $s(k)$ associated to Eqs.~(\ref{Eq:Warmup})~and~(\ref{Eq:Warmup_system}), respectively. In Fig.~\ref{Fig:Warmup_original} we show the real and imaginary parts of $s(k)$ for system~(\ref{Eq:Warmup}) with $\lambda=1$, where $S_{1}$ and $S_{2}$ denote each of the solutions $s(k)= \pm i k \sqrt{1-\lambda k^2}$.

\begin{figure}[h!]
\centering
\begin{minipage}{.5\textwidth}
  \centering
  \includegraphics[height=4.7cm,width=1\linewidth]{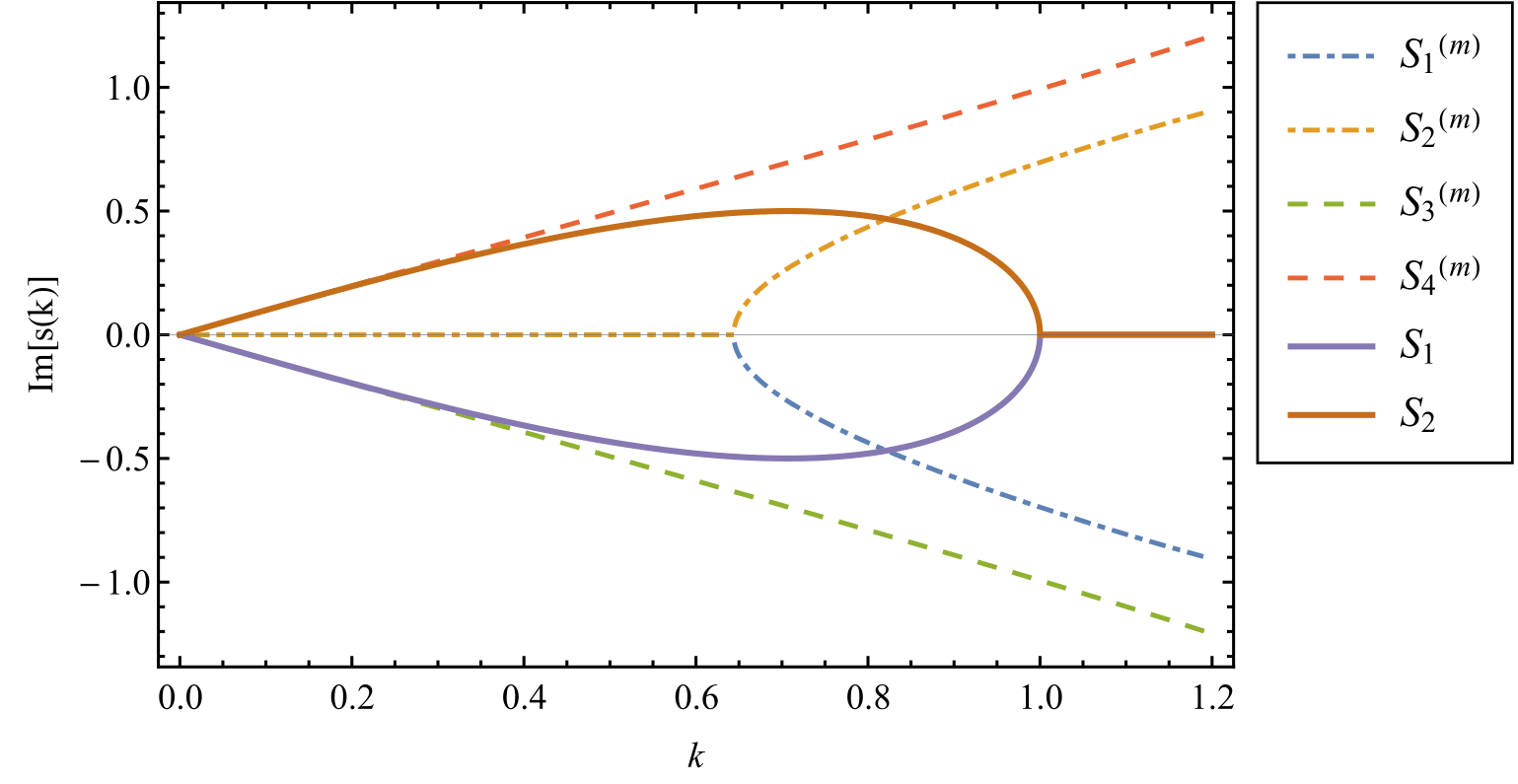}
\end{minipage}%
\begin{minipage}{.5\textwidth}
  \centering
  \includegraphics[height=4.7cm,width=1\linewidth]{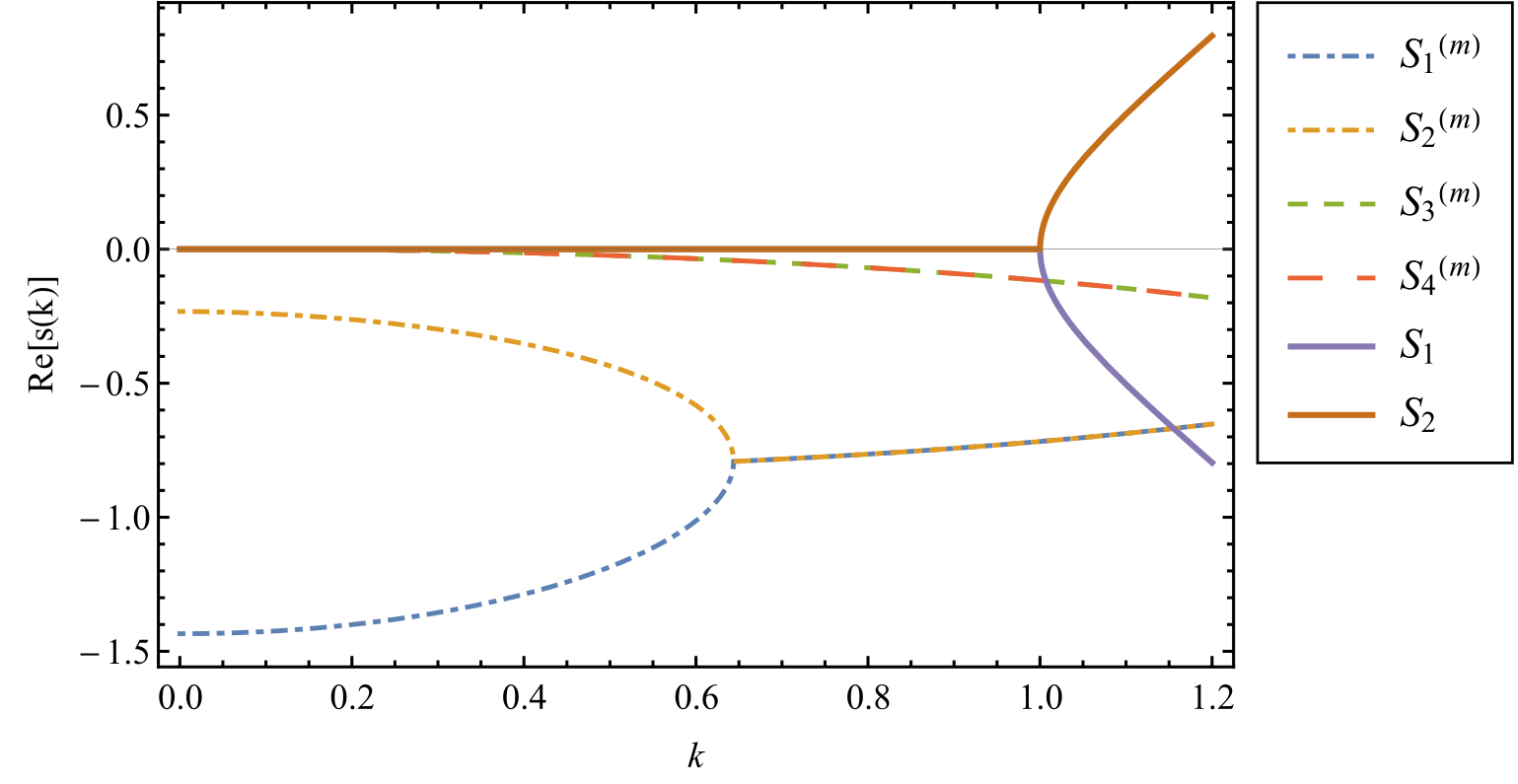}
\end{minipage}
\caption{\label{Fig:Warmup_modified} (Color online) Real and imaginary parts of $s(k)$ for the modified model Eq.~(\ref{Eq:Warmup_system}), with $\tau=5$, $\sigma=3$, and $\lambda=1$.}
\end{figure}

The appearance of blowing-up modes with $k>1$ is evident in the right panel of Fig.~\ref{Fig:Warmup_original}. Even if we argue that for values of $k$ sufficiently far away from the blow-up region the dispersion relation associated to this model could accurately describe physical modes, in a practical sense, the issues in the high frequency regime would not let us exploit the good aspects of the theory, and here is where the modified model~(\ref{Eq:Warmup_system}) steps in. Solving the quartic equation~(\ref{Eq:Quartic}), in Fig.~\ref{Fig:Warmup_modified} we show the behavior of its four solutions $s(k)$, where, as an example, we have set $\tau=5$, $\sigma=3$, and we have denoted the $i$-th solution by $S_{i}{}^{(m)}$. In particular, we observe that the magnitude of the real part of solutions $S_{3}{}^{(m)}$ and $S_{4}{}^{(m)}$ is smaller than that of solutions $S_{1}{}^{(m)}$ and $S_{2}{}^{(m)}$, thus corresponding to a slower decay. Moreover, this real part goes to zero as $k\rightarrow0$. Also, at sufficiently small $k$, the imaginary part of solutions $S_{3}{}^{(m)}$ and $S_{4}{}^{(m)}$ approaches that of $S_{1}$ and $S_{2}$, respectively, meaning that these low-$k$ modes present in the original system are still present in the modified model. These features of solutions $S_{3}{}^{(m)}$ and $S_{4}{}^{(m)}$ persist as we choose different values of $\tau$ and $\sigma$. The curves describing solutions $S_{1}{}^{(m)}$ and $S_{2}{}^{(m)}$ change their shapes when we vary the parameters, but in general we observe that $\Re[S_{1,2}{}^{(m)}] \leq \Re[S_{3,4}{}^{(m)}] \leq 0$.

The qualitative idea that we get from these observations is the following. The modified system~(\ref{Eq:Warmup_system}) 
captures solutions of the original model~(\ref{Eq:Warmup}) for sufficiently small values of $k$, which is good news provided that the inverse cascade hypothesis is correct. Moreover, high frequency blowing-up modes are no longer a problem. Even though ``spurious" modes also appear as a result of the modification [$S_{1}{}^{(m)}$ and $S_{2}{}^{(m)}$], these modes in general decay faster than those we identify with the original ones from 
Eq.~(\ref{Eq:Warmup}).

\subsubsection{On the dispersion relation for large $k$}

\begin{figure}[h!]
\centering
\begin{minipage}{.5\textwidth}
  \centering
  \includegraphics[height=4.7cm,width=1\linewidth]{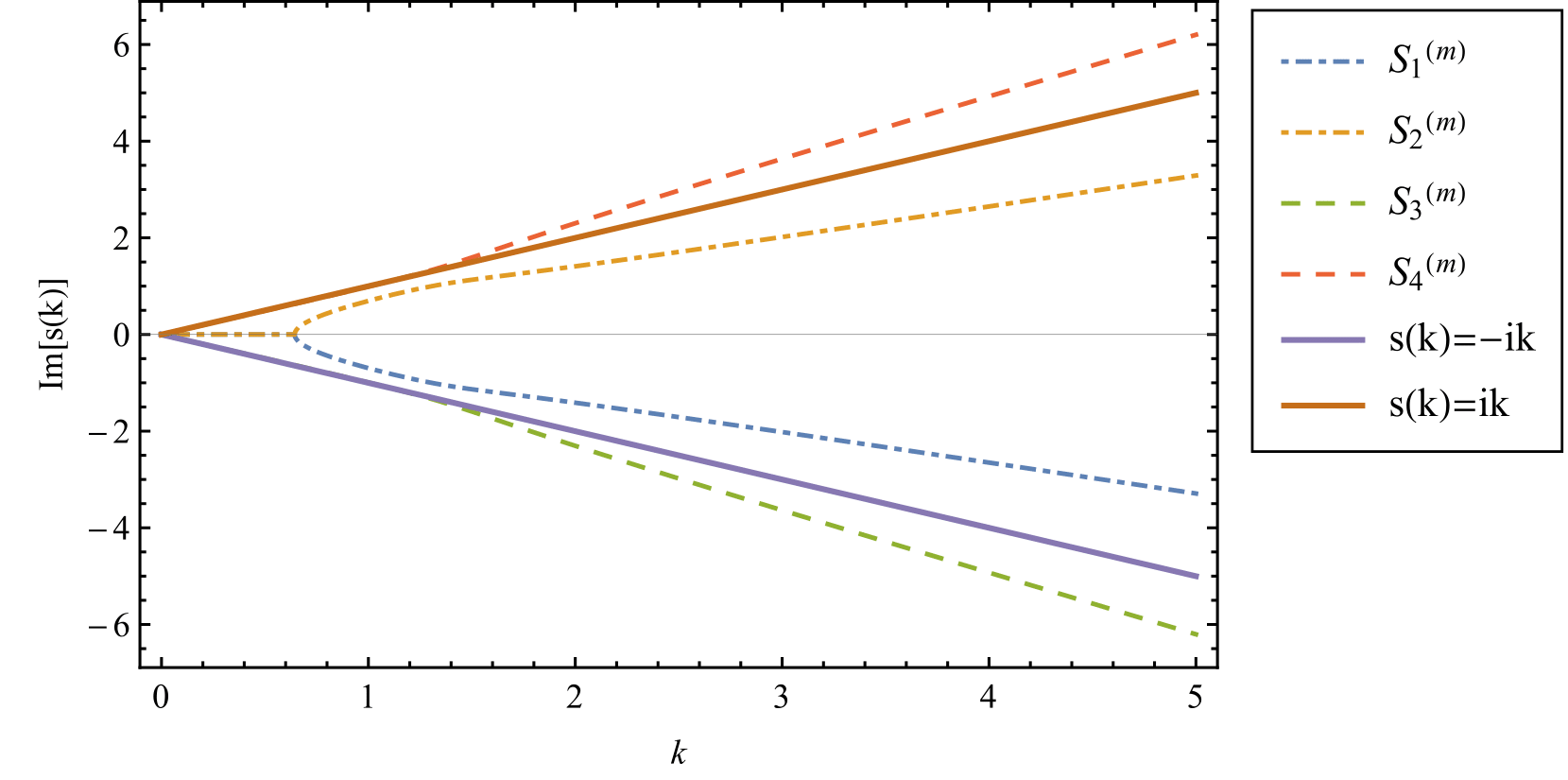}
\end{minipage}%
\begin{minipage}{.5\textwidth}
  \centering
  \includegraphics[height=4.7cm,width=1\linewidth]{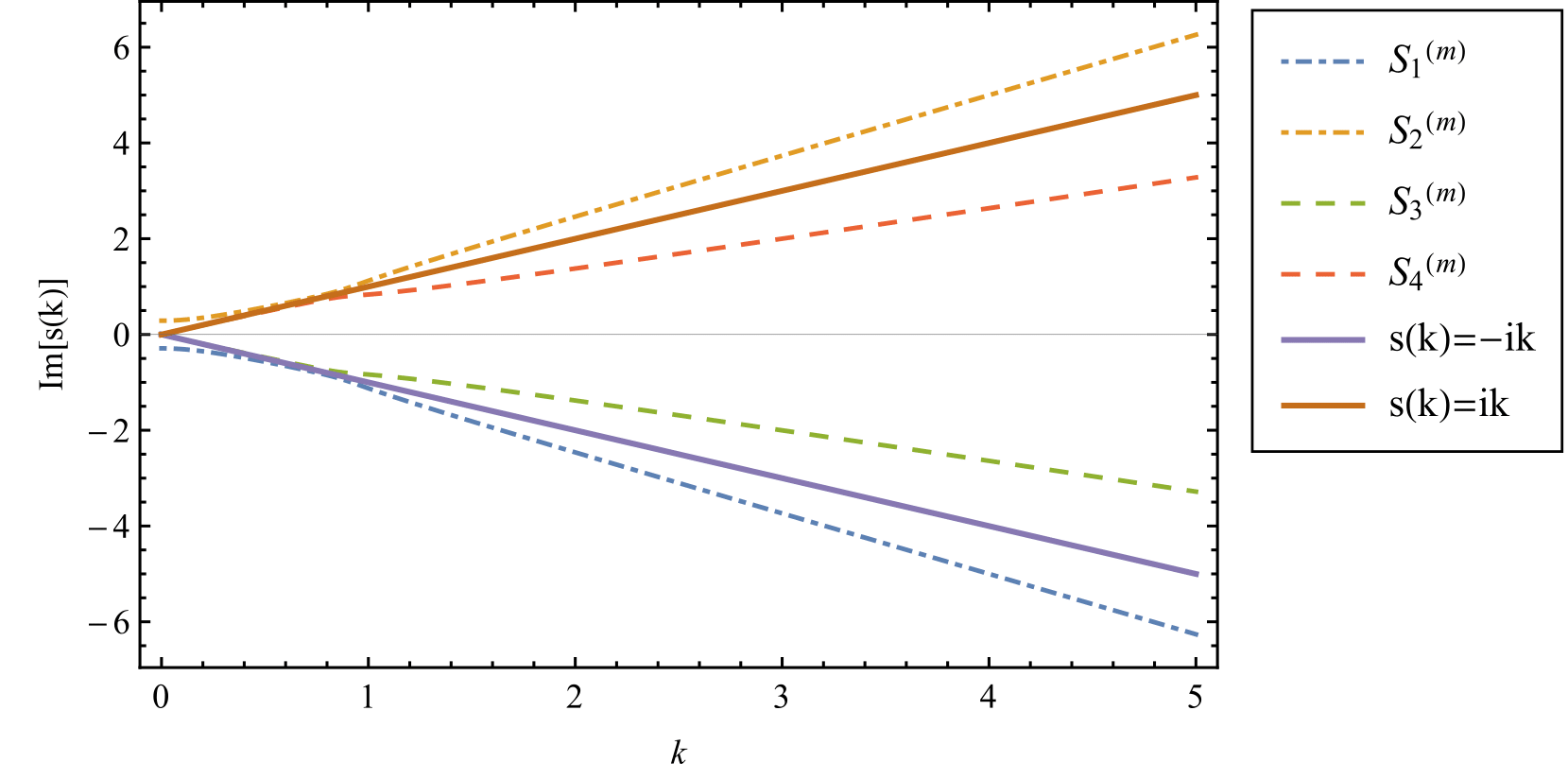}
\end{minipage}
\caption{\label{Fig:Velocidades} (Color online) Imaginary part of $s(k)$ for the modified model Eq.~(\ref{Eq:Warmup_system}) in comparison to $s(k)=\pm ik$, with $\lambda=1$, $\tau=5$, $\sigma=3$ (left panel) and $\lambda=1$, $\tau=2$, $\sigma=3$ (right panel).}
\end{figure}

To complete our discussion on the modified model represented by Eqs.~(\ref{Eq:Warmup_system}), we explore the behavior of the dispersion relation solutions $s(k)$ of Eq.~(\ref{Eq:Quartic}), as well as their phase velocity $s(k)/k$ in the high frequency regime. To this respect, numerical computations suggest that $s(k)/k$ asymptotes a constant complex value. This motivates an inspection of Eq.~(\ref{Eq:Quartic}) under the ansatz $s(k)=\alpha k$ with $\alpha\in\mathbb{C}$. Considering only leading terms, we find $\sigma(\alpha^2 + 1)^2 - \lambda = 0$, whose solutions are
\begin{equation}\label{Eq:Speeds}
\alpha_{(\pm,\pm)}=\pm\sqrt{-1\pm\sqrt{\frac{\lambda}{\sigma}}}.
\end{equation}
By virtue of the assumption $\sigma\geq\lambda$, these solutions are purely imaginary, as expected. Furthermore, as the magnitude of $\alpha_{(\pm,\pm)}$ corresponds to the phase speed $||s(k)/k||$, Eq.~(\ref{Eq:Speeds}) reveals that two of the solutions possess modes propagating at subluminal speeds, whereas the other two solutions posses superluminal---but bounded---modes, being the departure from the speed of light dependent on the ratio $\lambda/\sigma$. These expectations are confirmed in Fig.~\ref{Fig:Velocidades}. As a general feature, we have found that the superluminal modes correspond to the ``spurious" modes identified in Fig.~\ref{Fig:Warmup_modified} [$S_{1}{}^{(m)}$ and $S_{2}{}^{(m)}$], as long as $\sigma\gtrsim\tau$.

We have learned not only that the mode's velocity remains finite as $k$ grows large, but it can be controlled by the parameters $\sigma$ and $\tau$. Since the speed of all modes is bounded, the numerical solution of the problem would only require relatively straightforward techniques.

\subsection{Case $n=3$}

\begin{figure}[h!]
\centering
\begin{minipage}{.5\textwidth}
  \centering
  \includegraphics[height=4.7cm,width=0.95\linewidth]{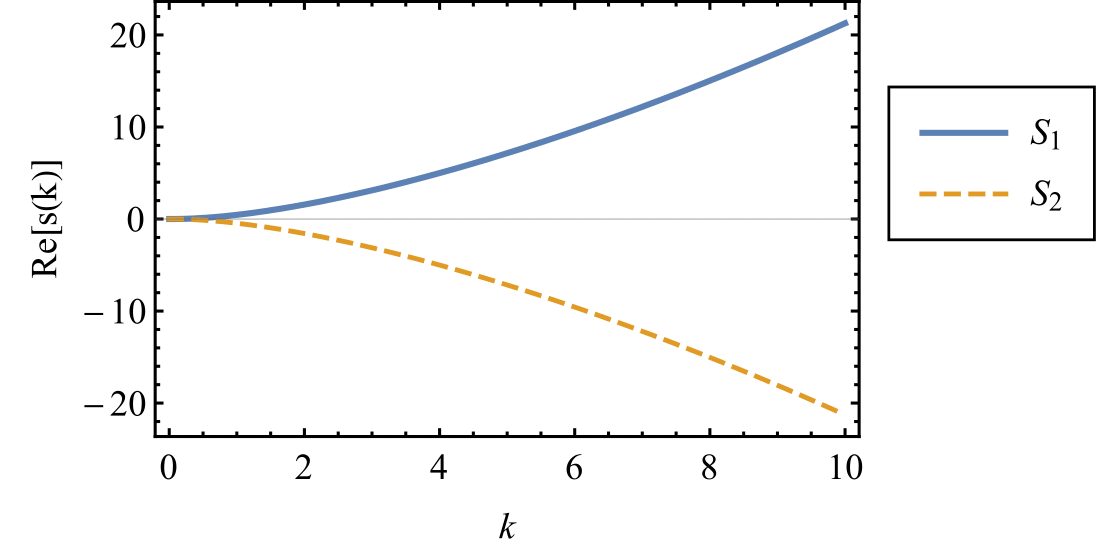}
\end{minipage}%
\begin{minipage}{.5\textwidth}
  \centering
  \includegraphics[height=4.7cm,width=0.95\linewidth]{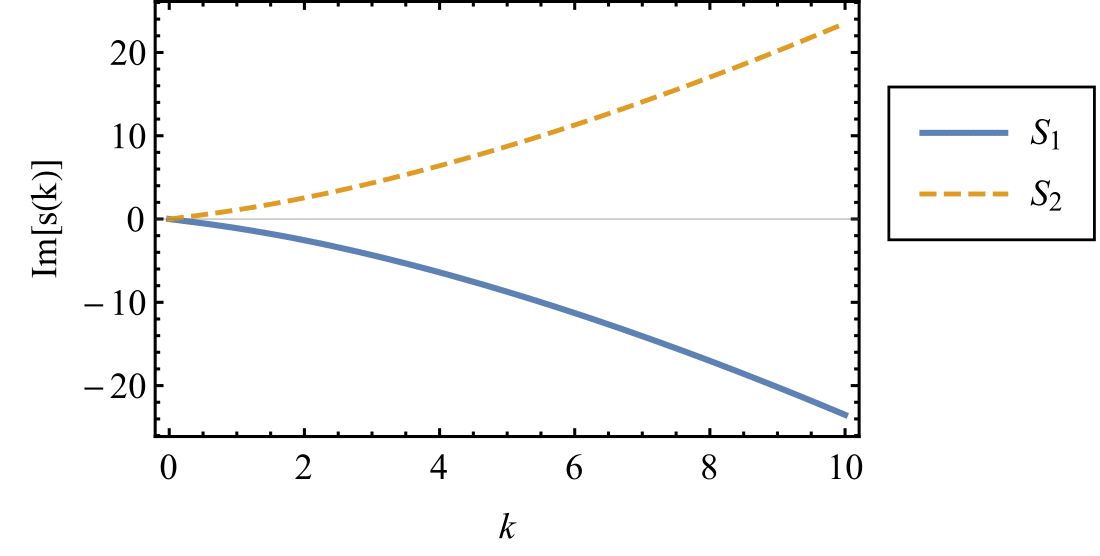}
\end{minipage}
\caption{\label{Fig:Warmup_n(3)} (Color online) Real and imaginary parts of the dispersion relation solutions $s(k)$ for the original, linearized model Eq.~(\ref{Eq:Warmup}) with $n=3$ and $\lambda=1$.}
\end{figure}

The case of $n=3$ has interesting properties which are not present in the case of $n=4$, as we shall show in the following discussion. First, the dispersion relation associated to Eq.~(\ref{Eq:Warmup}) is now given by 
\begin{equation}
s(k)~=~\pm ik\sqrt{1+\lambda ik},
\end{equation}
and because of the $i$ factor below the square root, we have blowing-up solutions for any $k$, as well as the existence of superluminal modes. (See Fig.\ref{Fig:Warmup_n(3)}.)

The more pathological behavior observed in this case, is due to the fact that the symmetry between the
order of time and space derivatives is broken in Eq.~(\ref{Eq:Warmup}). As we will show later,
there are indeed extensions to GR that contemplate such scenarios.

\begin{figure}[h!]
\centering
\begin{minipage}{.5\textwidth}
  \centering
  \includegraphics[height=4.8cm,width=1\linewidth]{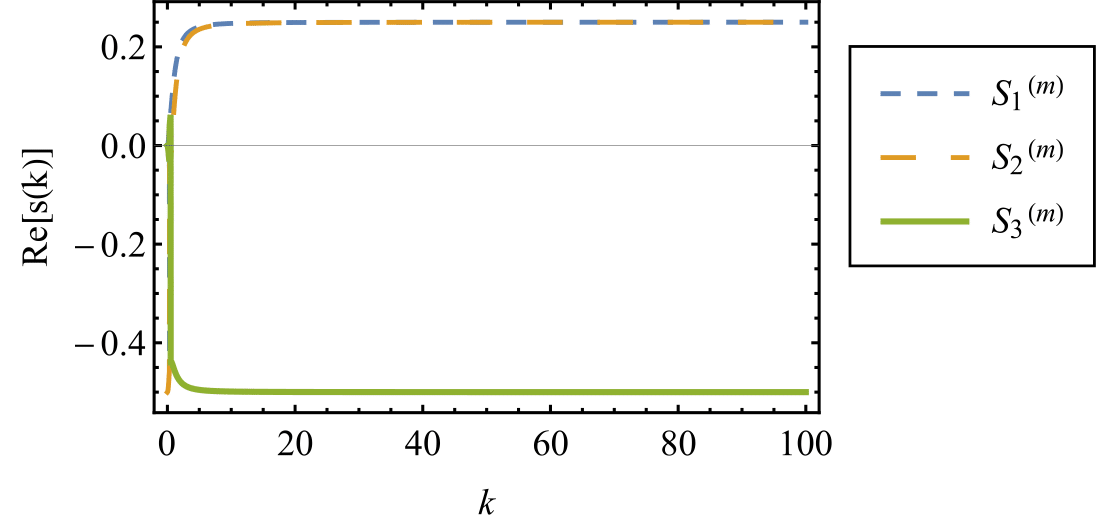}
\end{minipage}%
\begin{minipage}{.5\textwidth}
  \centering
  \includegraphics[height=4.8cm,width=1\linewidth]{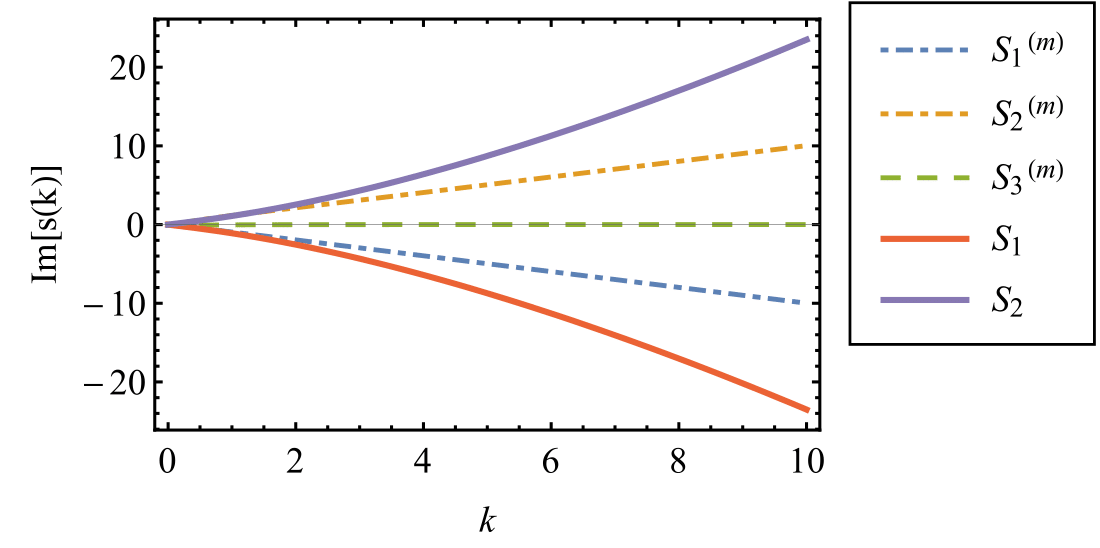}
\end{minipage}
\caption{\label{Fig:Warmup_modifiedn(3)} (Color online) Real and imaginary parts of $s(k)$ for the modified model Eq.~(\ref{Eq:Warmup_n(3)_system}), with $\tau=2$, $\sigma=2$, and $\lambda=1$.}
\end{figure}

As we did in the $n=4$ case, we introduce a modified system in an attempt to fix the solutions' pathologies. One possible modification is the following:
\begin{subequations}\label{Eq:Warmup_n(3)_system}
\begin{eqnarray}
\Box \phi &=& \lambda \partial_{x}\Pi, \\
\sigma\partial_{t}\partial_{x} \Pi  &=&\partial_{x}^{2} \phi-\tau \partial_t \Pi - \Pi. \label{Eq:Warmup_n(3)_system_b}
\end{eqnarray}
\end{subequations}
Notice that instead of introducing a term of the form $\sigma\Box\Pi$ in Eq.~(\ref{Eq:Warmup_n(3)_system_b}), we have instead employed a combination of one time derivative and one spatial derivative, which restores the symmetry related to space and time derivatives in the original equation. Using again the ansatz $\phi(t,x) = A\exp(s t + i k x)$, system (\ref{Eq:Warmup_n(3)_system}) implies
\begin{subequations}
\begin{eqnarray}
s(k)^2 + k^2 &=& \lambda\beta(s,k)ik, \\
\sigma\beta(s,k)iks &=& -k^2-\tau\beta(s,k)s(k)-\beta(s,k).
\end{eqnarray}
\end{subequations}
Eliminating $\beta$, we obtain the cubic equation for $s(k)$
\begin{eqnarray}\label{Eq:Qubic}
\Big{(}s(k)^2 + k^2\Big{)}\left(\sigma iks+{\tau}s(k)+1\right)=-{\lambda}ik^3,
\end{eqnarray}
whose solutions $S_{i}{}^{(m)}$ are presented in Fig.\ref{Fig:Warmup_modifiedn(3)}, as well as a comparison between the imaginary parts of these solutions and the ones corresponding to the original solutions $S_{i}$ of Eq.~(\ref{Eq:Warmup}).

\begin{figure}[H]
\centering
\begin{minipage}{.5\textwidth}
  \centering
  \includegraphics[height=4.8cm,width=1\linewidth]{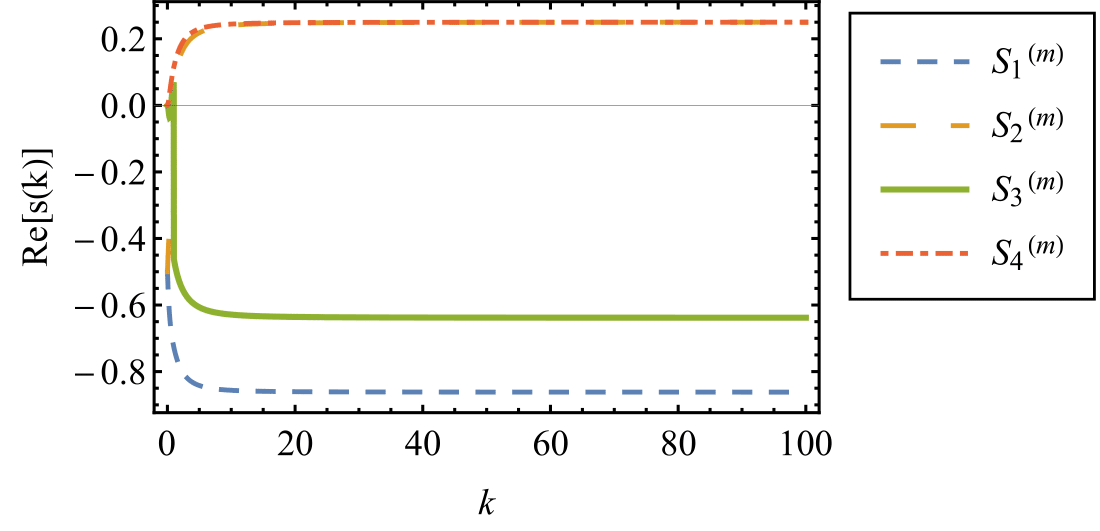}
\end{minipage}%
\begin{minipage}{.5\textwidth}
  \centering
  \includegraphics[height=4.8cm,width=1\linewidth]{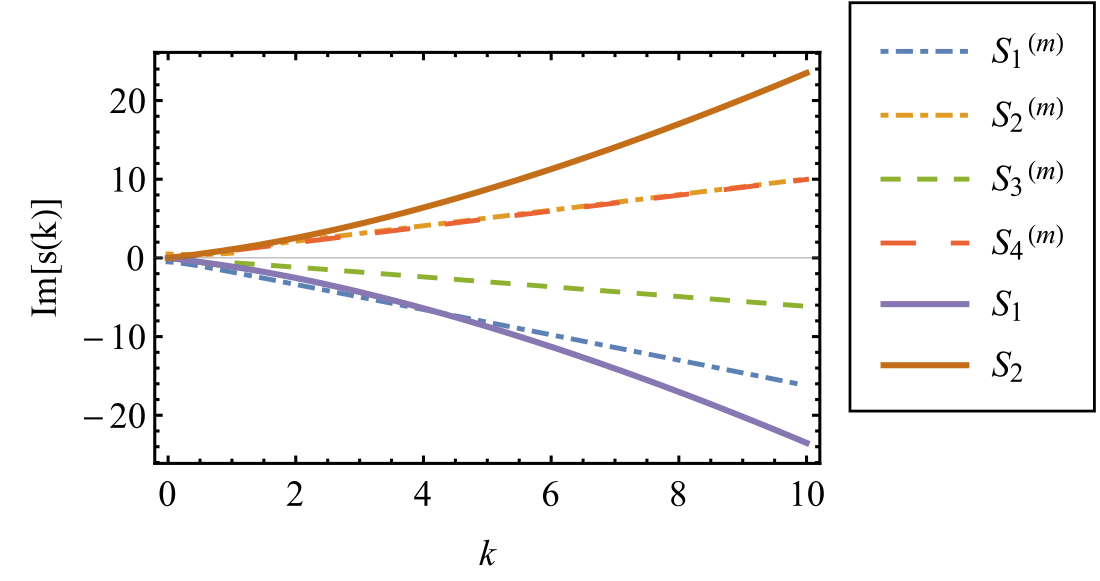}
\end{minipage}
\caption{\label{Fig:Warmup_modifiedn(3.2)} (Color online) Real and imaginary parts of $s(k)$ for the modified model Eq.~(\ref{Eq:Warmup_n(3)_system2}), with $\tau=2$, $\sigma=2$, $\lambda=1$, $\alpha$=2.}
\end{figure}

As can be inferred from the left panel of Fig.~\ref{Fig:Warmup_modifiedn(3)}, blowing-up modes still exist in this case. However, divergences are controlled as the real part of $s(k)$ is bounded from above for every $k$. Therefore, the problem is still well posed. Regarding the speed of mode propagation, we can see from the right panel of Fig.~\ref{Fig:Warmup_modifiedn(3)}, that the effect of the modification we introduce is to prevent the existence of superluminal modes.
A natural objection about the structure of Eq.~(\ref{Eq:Warmup_n(3)_system_b}) is the presence of the crossed derivative term which obscures the understanding of the evolution of the system. However, a simple modification can easily
address this issue.  Consider instead the addition of a term proportional to $\Box\Pi$, leading to the modified system
\begin{subequations}\label{Eq:Warmup_n(3)_system2}
\begin{eqnarray}
\Box \phi &=& \lambda \partial_{x}\Pi, \\
\alpha\Box\Pi+\sigma\partial_{t}\partial_{x} \Pi  &=&\partial_{x}^{2} \phi-\tau \partial_t \Pi - \Pi, \label{Eq:Warmup_n(3)_system2_b}
\end{eqnarray}
\end{subequations}
with $\alpha \in\mathbb{R}$. Now, both $\phi$ and $\Pi$ are determined by second-order in time equations. Importantly, the alternative system Eqs.~(\ref{Eq:Warmup_n(3)_system2}) also fixes the ill-posedness of Eq.~(\ref{Eq:Warmup}) in a qualitatively similar fashion as Eqs.~(\ref{Eq:Warmup_n(3)_system}). Repeating the analysis, in Fig.~\ref{Fig:Warmup_modifiedn(3.2)} we show that the high frequency dependent blow-up modes are controlled since the real part of $s(k)$ is bounded from above for every $k$, and propagation speeds are not superluminal.

Summarizing, in the original ill-posed model Eq.~(\ref{Eq:Warmup}) with $n=3$, blowing-up solutions and acausal propagation exist. We have alleviated these pathologies based on a prescription inspired in the Israel-Stewart approach. As a result, the blow up of solutions has been controlled in such a way that continuity in initial conditions is restored, and no more acausal mode propagation is present. We have also shown that successful modifications of the original problem are not unique [compare Eqs.~(\ref{Eq:Warmup_n(3)_system}) and~(\ref{Eq:Warmup_n(3)_system2})], underscoring how
versatile the strategy is.

\section{Examples in some extensions to General Relativity}
As we mentioned in the introduction, our goal is not to concentrate on a specific case of the many modifications that exist in the literature. Rather, our goal is to present a general approach that could be adopted in one's theory of choice. Having presented a thorough discussion of our toy models, we now illustrate the approach in a couple of extensions to GR: Noncommutative Geometry and Dynamical Chern-Simons gravity. Our choices are solely motivated by the disparate structure they possess and the challenge they present to our approach.  As discussed next, their intrinsic problems with respect to the question of well-posedness can be addressed by our strategy.

\subsection{Linearized Noncommutative Geometry}\label{Sec:Noncommutative}
This theory arises in the framework of Noncommutative Geometry (NCG), an effective theory that attempts to elucidate the nature of spacetime at high-energies scales~\cite{NCG:94,NCG2:08}. It is suspected that as energies approach the Planck scale, quantum gravity effects would manifest themselves as noncommutative properties of spacetime. The NCG framework suggests that for energies below, but close to the Planck scale, the spacetime can be considered as an ``almost" noncommutative manifold, more specifically, the product of a Riemannian spin manifold with a finite noncommutative space. Several of its aspects have attracted attention in the recent past, like the fact that it provides a geometrical derivation of the Standard Model; or that it has applications in high energy particle physics and early universe cosmology. As an example of the latter, Ref.~\cite{Sakellariadou:2011dk} explores constraints on the gravitational sector of the theory, and derives equations of motion for the metric $g_{ab}$ from a spectral action. To the best of our knowledge, this theory has not been explicitly shown to not admit a well posed initial value formulation. However, a cursory inspection of the resulting underlying equations hints this is very likely the case. To discuss a specific case, we consider a low energy, weak curvature regime, in which the nonminimal coupling between the background geometry and the Higgs field is neglected. In such a regime, the following field equations are obtained in four spacetime dimensions
\begin{equation}
R_{ab}-\frac{1}{2}g_{ab}R= \lambda\Big{(}2\nabla^c\nabla^dW_{acbd}+W_{acbd}R^{cd}\Big{)}+{\kappa}T^{\text{Matter}}_{ab} \label{ncg1},
\end{equation}
where $W_{acbd}$ is the Weyl tensor, $R_{ab}$ and $R$ are the Ricci tensor and scalar, respectively, $\nabla$ is the Levi-Civita connection, and $\lambda$, $\kappa^{-1} = 16\pi G_N$ are coupling constants. Here, we will focus in the vacuum case, so we take a vanishing stress-energy matter tensor $T^\text{Matter}_{ab}=0$. In this case, the traceless condition of the Weyl tensor implies $R=0$, so we are left with the equation
\begin{equation}
R_{ab}= \lambda \left(2\nabla^c\nabla^dW_{acbd}+W_{acbd}R^{cd}\right).
\end{equation}
We can rewrite the first term on the right-hand side of the equation above using the fact that, in four spacetime dimensions, in the case of vanishing Ricci scalar, the Weyl tensor can be written as 
\begin{equation}
W_{acbd}=R_{acbd}+\frac{1}{2}(g_{ad}R_{cb}+g_{cb}R_{ad}-g_{ab}R_{cd}-g_{cd}R_{ab}).
\end{equation}
Then, using Bianchi's Identity and its contracted version, which implies $2\nabla^aR_{ab}=\nabla_bR=0$, we obtain
\begin{equation}
\nabla^dW_{acbd} = \frac{1}{2}( \nabla_cR_{ab}-\nabla_aR_{bc}),
\end{equation}
and thus $2\nabla^c\nabla^dW_{acbd} = \left( {\Box}R_{ab}-\nabla_c\nabla_aR_{b}{^{c}} \right)$, so the equations of motion read
\begin{equation}
R_{ab}= \lambda \left( {\Box}R_{ab}-\nabla_c\nabla_aR_{b}{^{c}}+W_{acbd}R^{cd}\right).
\end{equation}

Let us now consider the linearized version of the theory around Minkowski spacetime. For this, we assume $g_{ab}=\eta_{ab}+h_{ab}$ with the metric perturbation $|h_{ab}| \ll |\eta_{ab}|$. We choose the Lorenz gauge, $\partial^{a}\bar{h}_{ab}=0$, where $\bar{h}_{ab}$ is the trace reversed perturbation, so we have $2R_{ab}=-{\Box}h_{ab}$, and we obtain the equations of motion up to first order in $h_{ab}$ 
\begin{equation}
{\Box}h_{ab}= \lambda \left( {\Box}{\Box}h_{ab}-\partial_c\partial_a{\Box}h_{b}{^c} \right).\label{ncg7}
\end{equation}
One can check, using the equation above for the components of $h_{ab}$  and for its trace, that it is consistent to use the residual gauge freedom from the Lorentz gauge to set $h_{0i}=0$ and $h=0$. In this case, the Lorentz gauge condition becomes $\partial^{a}{h}_{ab}=0$, and the equations of motion~(\ref{ncg7}) reduce to
\begin{equation}
{\Box}h_{ij}= \lambda{\Box}{\Box}h_{ij}.\label{ncg8}
\end{equation}
Now, we study Eq.~(\ref{ncg8}) from the same perspective as Eq.~(\ref{Eq:Warmup}). We first consider the ansatz $h_{ij}(t,\vec{x})=A\epsilon_{ij}e^{st+i\vec{k}.\vec{x}}$, where $\epsilon_{ij}$ is a symmetric traceless spatial tensor orthogonal to $\vec{k}$, with $\vec{x}$, $\vec{k} \in \mathbb{R}^3$, $A\in\mathbb{R}$. With this ansatz, Eq.~(\ref{ncg8}) implies the algebraic equation for the dispersion relation $s(k)$
\begin{equation}
\lambda \left(s(k)^2 + |k|^2 \right)^2 - \left( s(k)^2+|k|^2 \right) = 0,
\end{equation}
whose solutions are
\begin{equation}
s(k) = \pm i|k| \quad \mbox{~and~} \quad s(k) = \pm\sqrt{ \frac{1}{\lambda}-|k|^2}.
\end{equation}
The first pair is a sensible set of solutions propagating at the speed of light, whereas the second pair gives rise to blow-up modes if $\lambda>0$ and $|k|$ is sufficiently small. Notice that this feature is different to the one that we were tackling so far, namely ill-posedness evidenced by the existence of high frequency blowing-up modes. However, our strategy is not exclusively applicable to fix the uncontrolled growth of high frequency modes. Moreover, in this case there is no signature of acausality because the modes' velocity is bounded in any frequency regime. Notice also that the solutions $s(k)$ behave well if we assume $\lambda<0$, which is precisely the case being explored in the literature~\cite{Sakellariadou:2011dk}. Nevertheless, the blowing-up of low frequency modes is indeed an issue {\it per se}, and thus we employ this model as a convenient example of the application of the Israel-Stewart line of thought in a gravity context. In order to show this, let us consider the system
\begin{subequations}\label{Eq:ncm_system}
\begin{eqnarray}
{\Box}h_{ij}&=& \lambda{\Box}\Pi_{ij}, \\
\Pi_{ij}&=& {\Box}h_{ij}-\tau\partial_t\Pi_{ij}-\sigma{\Box}\Pi_{ij}.
\end{eqnarray}
\end{subequations}
Taking the usual ansatz for the new dynamic field $\Pi_{ij}(t,\vec{x})=A\beta(s,\vec{k})\epsilon_{ij}e^{st+i\vec{k}}$, Eqs.~(\ref{Eq:ncm_system}) imply that the associated dispersion relation $s(k)$ satisfies
\begin{equation}
 \lambda \left(s(\vec{k})^2 + |k|^2 \right)^2 - \left( s(\vec{k})^2 + |k|^2 \right) \left[ \sigma \left( s(\vec{k})^2+|k|^2 \right) + \tau s(\vec{k}) + 1 \right] = 0,
\end{equation}
with solutions
\begin{equation}
s(\vec{k}) = \pm i|k| \quad \mbox{~and~} \quad s(\vec{k}) = -\frac{{\tau}}{2(\sigma-\lambda)}\pm\sqrt{\left( \frac{{\tau}}{2(\sigma-\lambda)} \right)^2-\frac{1}{(\sigma-\lambda)}-|k|^2}.
\end{equation}
The first set of solutions exhibits the same modes travelling at the speed of light as the original Eq.~(\ref{ncg8}). The second set is such that $\Re[s(\vec{k})]<0$ for all ${\vec{k}}$, provided $\sigma>\lambda$ and $\tau>0$. It is also easy to check that in the case of $\sigma=\lambda$, $\Re[s(\vec{k})]<0$ for all ${\vec{k}}$, provided $\tau>0$. Still in this case, the dispersion velocity keeps bounded in any frequency regime. In particular, it approaches the speed of light as $k$ grows arbitrarily large. Notice that the phase velocity's magnitude $\|s(\vec{k})/k\|$ asymptotes the speed of light from below as $k$ growths arbitrarily large.

In summary, we have considered a rather simplified model of linearized gravity in the context of Noncommutative Geometry. We have found the existence of blowing-up modes in the low frequency regime of solutions of the equation governing the propagation of metric perturbations. Next, we have suitably modified this equation with our approach, and we have shown that solutions do not have blowing-up modes anymore.

\subsection{Linearized Chern-Simons gravity}\label{Sec:CS}
Chern-Simons (CS) gravity is an effective extension of General Relativity~\cite{Jackiw_and_Pi03,Alexander_Yunes09}. It is mainly motivated by parity-violating requirements from particle physics, and can be derived from different approaches, in particular via a low energy limit of string theory~\cite{Polchinski98}. As in the previous example, well-posedness of CS gravity is formally an open question and, even when in certain specialized cases the theory has been proven to be well posed as a Dirichlet boundary value problem~\cite{Grumiller_etal08}, its structure in general situations again hints strongly that it will not lead to well posed problems in general. Indeed, the presence of higher order derivatives in the equations of motion can lead to the emergence of unstable modes, unless CS is regarded as an effective field theory for which only linear terms are kept in a perturbative expansion around GR~\cite{Delsate_etal15}. The authors of the same work conclude that, in the full, nonperturbative case, there are strong suspicions of ill-posedness of the CS evolution system. For our current purposes, the presence of higher derivative terms in CS gravity constitutes a suitable scenario to apply the ideas presented throughout this work. Let us consider a 4-dimensional spacetime $({\cal M},g_{ab})$ with Levi-Civita connection $\nabla$. The CS action is given by
\begin{equation}\label{Eq:CS_action}
S = \int_{\cal M} d^4x \sqrt{-g} \left\{ \kappa R + \frac{\alpha}{4} \vartheta~{}^*R R - \frac{\beta}{2} \left[ \nabla^a \vartheta \nabla_a \vartheta + 2V(\vartheta) \right] + {\cal L}_\text{Matter} \right\},
\end{equation}
where $\kappa^{-1} = 16\pi G_N$, $\alpha$ and $\beta$ are dimensional coupling constants, $R$ is the Ricci scalar, $g$ is the determinant of the metric tensor $g_{ab}$, and $\vartheta$ is a real, scalar coupling field. The Lagrangian density ${\cal L}_\text{Matter}$ corresponds to matter fields independent of $\vartheta$. The Pontryagin density is defined by
\begin{equation}
{}^*RR := -\frac{1}{2}\tensor{\epsilon}{^{cd}_{ef}}\tensor{R}{^{abef}}\tensor{R}{_{abcd}},
\end{equation}
where $\epsilon\indices{^{cdef}}$ is the 4-dimensional Levi-Civita tensor. Notice that in the case where $\alpha = 0$ and $\beta = 1$, CS reduces to GR with the scalar field $\vartheta$ minimally coupled to the gravity sector.

The extrema of action~(\ref{Eq:CS_action}) with respect to $g_{ab}$ and $\vartheta$, respectively lead to the field equations and scalar field equation of motion
\begin{subequations}\label{Eq:CS_field_eqns}
\begin{eqnarray}
G_{ab} + \frac{\alpha}{\kappa} C_{ab} &=& \frac{1}{2\kappa} T_{ab}, \\
\Box \vartheta + \frac{\alpha}{4\beta} {}^*RR &=& \frac{dV}{d\vartheta},\label{Eq:CS_scalar_eqn}
\end{eqnarray}
\end{subequations}
where $G_{ab} = R_{ab} - \frac{1}{2}g_{ab}R$ is the Einstein tensor, $C_{ab} := \nabla_c \vartheta \epsilon \indices{^{cd}_{e(a}} \nabla^e R\indices{_{b)}_d} + \nabla^c \nabla^d \vartheta~{}^*R\indices{_{d(ab)c}}$ is the so called $C$-tensor, and $T_{ab}$ is the total stress-energy tensor given by
\begin{equation}
T_{ab} = -\frac{2}{\sqrt{-g}} \left( \frac{\delta {\cal L}_\text{Matter}}{\delta g_{ab}}  + \frac{\delta {\cal L}_{\vartheta}}{\delta g_{ab}} \right),
\end{equation}
where ${\cal L}_{\vartheta} = - (\beta/2) \left[ \nabla^a \vartheta \nabla_a \vartheta + 2V(\vartheta) \right]$.

In the following, we restrict to the vacuum case, as well as vanishing scalar field potential, meaning ${\cal L}_\text{Matter} = 0$ and $V(\vartheta) = 0$. In this case, the field equations in trace-reversed form, and the scalar field equation of motion are
\begin{eqnarray}
R_{ab}+\frac{\alpha}{\kappa}C_{ab} &=& \frac{1}{2\kappa}\left( T_{ab}-\frac{1}{2}g_{ab}T \right), \label{Eq:Eq_Tensor}\\
\Box\vartheta + \frac{\alpha}{4\beta}{}^{*}RR &=& 0, \label{Eq:Eq_Scalar}
\end{eqnarray}
with
\begin{equation}
T_{ab} = \beta \left( \nabla_a \vartheta \nabla_b \vartheta - \frac{1}{2}g_{ab} \nabla^c \vartheta \nabla_c  \vartheta \right) \quad \mbox{and} \quad T = g^{cd}T_{cd}.
\end{equation}

A subclass of CS gravity assumes a nondynamical scalar field $\vartheta$. This framework is set by choosing the coupling constant $\beta = 0$ at the level of the action~(\ref{Eq:CS_action}), further implying the Pontryagin constraint ${}^*RR=0$, which is rather restrictive in the allowed types of spacetime solutions. An extended discussion on the nondynamical formulation can be found in~\cite{Alexander_Yunes09}. As opposed to the nondynamical case, the more generic {\it dynamical Chern-Simons} (dCS) framework, in which $\vartheta$ is allowed to evolve according to Eq.~(\ref{Eq:CS_scalar_eqn}), is less restrictive and thus has the potential to be physically more relevant. For this reason, in this work we focus on the dCS extension of GR.

\subsubsection{Linearization of dCS gravity}\label{Sec:dCS_Lin}

We now proceed to study the linearized dynamical Chern-Simons theory. For this, we consider perturbations of both the metric and the scalar field,
\begin{equation}
g_{ab}=\bar{g}_{ab}+h_{ab}, \label{Eq:Pert_g}
\end{equation}
\begin{equation}
\vartheta=\bar{\vartheta}+\delta\vartheta, \label{Eq:Pert_theta}
\end{equation}
where $\bar{g}_{ab}$ and $\bar{\vartheta}$ constitute a background solution of Eqs.~(\ref{Eq:Eq_Tensor})-(\ref{Eq:Eq_Scalar}).\footnote{The same problem was treated in Refs.~\cite{dGfPnY10,aDkYnY14} with a decomposition of background and perturbed quantities suitable to their applications. In Ref.~\cite{hMtS14}, linear perturbations at the level of the action were performed, which unveiled the presence of ghost degrees of freedom in the case of $\bar{\vartheta} \ne 0$.}
Our analysis will be focused on high frequency solutions of the perturbed equations of motion, aiming to spot 
pathologies due to higher derivatives in this particular regime. Our approach to high frequencies assumes the geometric optics approximation, which we adopt by considering the following ansatz for the metric and scalar field perturbations,
\begin{eqnarray}
h_{ab} &=& A_{ab}(t,x^j)\exp\left[ i\phi(t,x^k)/\epsilon_{\phi} \right], \label{Eq:ansatz_h}\\
\delta\vartheta &=& B(t,x^j)\exp\left[ i\phi(t,x^k)/\epsilon_{\phi} \right],\label{Eq:ansatz_delta}
\end{eqnarray}
where the requirement that the phase $\phi$ varies much faster than the amplitudes ($A_{ab},B)$ is achieved by taking the parameter $\epsilon_{\phi}\ll 1$. Notice that we are considering steady-state solutions since the metric and scalar perturbations share the same phase function.
The way we discriminate terms in our approximation is the following. Terms including derivatives of the phase function, $\nabla_{a}\phi$, will be accompanied by factors of $\epsilon_{\phi}^{-1}$. From the structure of the equations of motion, it is easy to see that we will have four kinds of terms, having from zero to three derivatives of the phase function. The third derivative term comes from the portion of the equations of motion that we are most interested in study, so we will keep it. As we want to stay as close to leading order as possible, we will also keep second derivative terms, but we will drop the lower derivative ones. Now, since the linearized energy momentum tensor and its trace can only have at most one derivative of $\phi$ at linear order in the perturbation, we can ignore the right-hand side of Eq.~(\ref{Eq:Eq_Tensor}) and linearize instead
\begin{equation}
R_{ab}+\frac{\alpha}{\kappa}C_{ab}=0. \label{Eq:Eq_Tensor2}
\end{equation}
Dropping terms with fewer than two derivatives of the perturbation, we obtain\footnote{Our perturbative calculations were all checked using the free package {\it xPert}~\cite{dBjmMGgaMM09,xPert} for {\it Mathematica}~\cite{mathematica}.}
\begin{eqnarray}
\left( R_{ab}+\frac{\alpha}{\kappa}C_{ab} \right)^{(1)} = &-&\frac{1}{2}\Big{(}\bar{\nabla}_{a}\bar{\nabla}_{b}h+\bar{\nabla}_{c}\bar{\nabla}^{c}h_{ab}\Big{)} + \bar{\nabla}^{c}\bar{\nabla}_{(a}h_{b)c} \nonumber \\
&+&\frac{\alpha}{\kappa}{}^{*}\bar{R}_{c(ab)d}\bar{\nabla}^{d}\bar{\nabla}^{c}\delta\vartheta-\frac{\alpha}{2\kappa}\left(\bar{\nabla}^{d}\bar{\nabla}^{c}\bar{\vartheta}\right)\bar{\epsilon}_{(a|def|}\bar{\nabla}_{b)}\bar{\nabla}^fh^{e}_{~c} \nonumber \\
&+&\frac{\alpha}{2\kappa}\left(\bar{\nabla}^{d}\bar{\nabla}^{c}\bar{\vartheta}\right)\bar{\epsilon}_{(a|def}\bar{\nabla}_{c|}\bar{\nabla}^fh^{e}_{~b)} \nonumber \\
&-&\frac{\alpha}{2\kappa}\left(\bar{\nabla}^{c}\bar{\vartheta}\right)\bar{\epsilon}_{(a|cef}\bar{\nabla}^{f}\bar{\nabla}_{d|}\bar{\nabla}_{b)}h^{de}\nonumber \\
&-&\frac{\alpha}{2\kappa}\left(\bar{\nabla}^{c}\bar{\vartheta}\right)\bar{\epsilon}_{(a|cef}\bar{\nabla}^{f}\bar{\nabla}_{d|}\bar{\nabla}^{e}h_{~b)}^{d}\nonumber \\
&+&\frac{\alpha}{2\kappa}\left(\bar{\nabla}^{c}\bar{\vartheta}\right)\bar{\epsilon}_{(a|cdf}\bar{\nabla}^{f}\bar{\nabla}_{e|}\bar{\nabla}^{e}h_{~b)}^{d}\nonumber \\
&+&\frac{\alpha}{2\kappa}\left(\bar{\nabla}^{c}\bar{\vartheta}\right)\bar{\epsilon}_{(a|cef|}\bar{\nabla}^{f}\bar{\nabla}^{e}\bar{\nabla}_{b)}h +\mathcal{O}\left(\epsilon_{\phi}^{-1}\right), \label{Eq_Tensorlin1}
\end{eqnarray}
where symbols with an over bar correspond to background functions and operators, and $h$ is the trace of the metric perturbation. Towards linearization of Eq.~(\ref{Eq:Eq_Scalar}), on the one hand we have
\begin{eqnarray}
\Box\vartheta^{(1)}&=&\Box\bar{\vartheta}+\bar{\Box}\delta\vartheta-h^{ab}{\bar{\nabla}}_a{\bar{\nabla}}_b\bar{\vartheta} \nonumber \\
&-&\frac{1}{2}\bar{g}^{ab}\Big{[}\bar{g}^{cd}(\partial_{a}h_{bd}+\partial_{b}h_{da}-\partial_{d}h_{ab}) \nonumber \\
&-&h^{cd}(\partial_{a}\bar{g}_{bd}+\partial_{b}\bar{g}_{da}-\partial_{d}\bar{g}_{ab})\Big{]}\partial_{c}\bar{\vartheta} \nonumber \\
&=&\bar{\Box}\delta\vartheta+\mathcal{O}\left(\epsilon_{\phi}^{-1}\right), \label{Eq:linscal_left}
\end{eqnarray}
where in the last line we have only kept the term with second derivatives of the perturbation; and on the other hand we have
\begin{equation}
\left({}^{*}RR\right)^{(1)} = 2\bar{R}_{abcd}\bar{\epsilon}^{cdef}\bar{\nabla}^{a}\bar{\nabla}_eh^{b}_{~f}+\mathcal{O}\left(\epsilon_{\phi}^{-1}\right).
\end{equation}
Therefore, the linearized version of Eq.~(\ref{Eq:Eq_Scalar}) is
\begin{equation}\label{Eq:linearized_scalar_eqn}
\bar{\Box}\delta\vartheta + \frac{\alpha}{2\beta}\bar{R}_{abcd}\bar{\epsilon}^{cdef}\bar{\nabla}^{a}\bar{\nabla}_eh^{b}_{~f} = 0.
\end{equation}
In order to express the perturbed equations in terms of the wave vector $k_a := \nabla_{a}\phi$, we take into account the following considerations:
\begin{itemize}
\item[(i)] Quantities of the form $\bar{\nabla}_{a}\bar{\nabla}_bh_{cd}$ ( or $\bar{\nabla}_{a}\bar{\nabla}_b\delta\vartheta$)  can be replaced by  $\partial_{a}\partial_bh_{cd}=k_{a}k_bh_{cd}$ ($\partial_{a}\partial_b\delta\vartheta=k_{a}k_b\delta\vartheta$) since these are the only terms of order $\epsilon_{\phi}^{-2}$.
\item[(ii)] Quantities of the form $\bar{\nabla}_{a}\bar{\nabla}_b\bar{\nabla}_{c}h_{de}$ have to be examined with more detail since they contain terms of order $\epsilon_{\phi}^{-2}$ and $\epsilon_{\phi}^{-3}$. This means that background Christoffel symbols will appear in the $\epsilon_{\phi}$ expansion.
\item[(iii)] Since the tensor $C_{ab}$ is traceless, it is consistent to choose the transverse traceless gauge for the metric perturbation, thus terms containing $h$ or $k^{a}h_{ab}$ will vanish.
\item[(iv)] In order to isolate the metric perturbation $h_{ab}$ from the scalar field perturbation $\delta \vartheta$, we multiply Eq.~(\ref{Eq_Tensorlin1}) by $k_n k^n$, and we replace the resulting term $\bar{\Box}\delta\vartheta$ using Eq.~(\ref{Eq:linearized_scalar_eqn}).
\end{itemize}

With these considerations, the equation for the metric perturbation takes the form
\begin{eqnarray}
0 &=& -\frac{1}{2}(k_ck^c)^2 h_{ab} \nonumber \\
&+& \Bigg{[}     B^{(1)}_{(a|fel}k^fk^e(k_ck^c)+B^{(2)}_{(a|l}(k_ck^c)^2 +B^{(3)}_{(a|fl}k^f(k_ck^c)^2 \Bigg{]} h_{b)}{}^{l} \nonumber \\
&+&\Bigg{[}       B^{(4)}_{abfedl}k^fk^e(k_ck^c)+B^{(5)}_{abdl}(k_ck^c)^2+ B^{(6)}_{abfespdl}k^fk^ek^sk^p\Bigg{]}h^{dl},\label{Eq_Tensorlin2}
\end{eqnarray}
where the tensors $B^{(i)}$ are combinations of the background quantities. Being explicit:
\begin{eqnarray}
B^{(1)}_{afel}&=&\frac{\alpha}{2\kappa}\Bigg{[} \left(\bar{\nabla}^d\bar{\nabla}_e\bar{\vartheta}\right)\bar{\epsilon}_{adlf}  -\left(\bar{\nabla}^c\bar{\vartheta}\right)\bar{\epsilon}_{ac}{}^{n}{}_{f}\bar{\Gamma}^{d}_{nl}\bar{g}_{de} \nonumber \\
&-& \left(\bar{\nabla}^c\bar{\vartheta}\right)\bar{\epsilon}_{acef}\bar{\Gamma}^{d}_{dl} 
+\left(\bar{\nabla}^c\bar{\vartheta}\right)\bar{\epsilon}_{acnf}\bar{\Gamma}^{n}_{le}
+\left(\bar{\nabla}^c\bar{\vartheta}\right)\bar{\epsilon}_{aclf}\bar{\Gamma}^{n}_{ne}\Bigg{]}, \nonumber \\
B^{(2)}_{al}&=& \frac{\alpha}{2\kappa}\left(\bar{\nabla}^c\bar{\vartheta}\right)\bar{\epsilon}_{acd}{}^{f}\bar{\Gamma}^{d}_{fl},\nonumber \\
B^{(3)}_{afl}&=& \frac{\alpha}{2\kappa}\left(\bar{\nabla}^c\bar{\vartheta}\right)\bar{\epsilon}_{aclf}, \nonumber \\
B^{(4)}_{abfeld}&=&\frac{\alpha}{2\kappa}\Bigg{\{}\Big{[} -\left(\bar{\nabla}^c\bar{\nabla}_d\bar{\vartheta}\right)\bar{\epsilon}_{(a|clf|} -\left(\bar{\nabla}^c\bar{\vartheta}\right)\bar{\epsilon}_{(a|cdf}\bar{\Gamma}^{n}_{nl|} -\left(\bar{\nabla}^c\bar{\vartheta}\right)\bar{\epsilon}_{(a|cnf}\bar{\Gamma}^{n}_{dl|}\Big{]}\bar{g}_{b)e} \nonumber \\
&+&\Big{[}-\left(\bar{\nabla}^c\bar{\vartheta}\right)\bar{\epsilon}_{(a|cdf|}\bar{\Gamma}^{n}_{b)l}\bar{g}_{ne}+\left(\bar{\nabla}^c\bar{\vartheta}\right)\bar{\epsilon}_{(a|clf|}\bar{\Gamma}^{n}_{b)d}\bar{g}_{ne} \nonumber \\
&+&\left(\bar{\nabla}^c\bar{\vartheta}\right)\bar{\epsilon}_{(a|cef|}\bar{\Gamma}^{n}_{b)d}\bar{g}_{nl}-2\left(\bar{\nabla}^c\bar{\vartheta}\right)\bar{\epsilon}_{(a|cdf|}\bar{\Gamma}^{n}_{b)e}\bar{g}_{nl}\Big{]}\Bigg{\}}, \nonumber \\
B^{(5)}_{abdl}&=&-\left(\bar{\nabla}^c\bar{\vartheta}\right)\bar{\epsilon}_{(a|cd|}{}^{f}\bar{\Gamma}^{n}_{b)f}\bar{g}_{nl}, \nonumber \\
B^{(6)}_{abfespdl}&=&-\frac{\alpha^2}{\beta\kappa}{}^{*}\bar{R}_{p(ab)s}\bar{R}_{fdcn}\bar{\epsilon}^{cn}{}_{el}.\nonumber 
\end{eqnarray}
Clearly, Eq.~(\ref{Eq_Tensorlin2}) is rather convoluted, as it mixes components of the metric perturbation contracted with tensors that depend on unspecified background quantities. However, the scope of this work does not reach details of any particular theory. We aim to discuss on our prescription to alleviate possible ill behavior of solutions in the high frequency regime. Thus we leave the background unspecified\footnote{In fact, the assumption of specific backgrounds
might even yield misleading answers as to the general behavior.}, and concentrate instead in the structure of the perturbation Eq.~(\ref{Eq_Tensorlin2}).

\subsubsection{dCS toy model}\label{Sec:dCS_Toy}

Having in mind the observation in item~(i), a $1+1$ dimensional toy model can be motivated by the structure of Eq.~(\ref{Eq_Tensorlin2}), namely
\begin{equation}\label{Eq:Model2}
\Box\Box\varphi+a\partial_{x}^{2}\Box{\varphi}+b\partial^{4}_{x}\varphi+\lambda\partial_{x}\Box\Box\varphi=0,
\end{equation}
where $\varphi$ plays the role of the perturbation, and $(a,b,\lambda)$ are real coefficients. From this simple model we aim to show how the highest derivative term, the one with coefficient $\lambda$, can spoil well-posedness and how could we fix such a problem. For this purpose, we take the following strategy. First, we consider Eq.~(\ref{Eq:Model2}) with $\lambda=0$ and choose parameters $(a,b)$ such that the system possesses no blow-up modes. Then we will show how the addition of the high derivative term can spoil the original system. Finally, we will apply our method to fix the problematic equation.

\begin{figure}[h!]
\centering
\begin{minipage}{.5\textwidth}
  \centering
  \includegraphics[height=4.6cm,width=1\linewidth]{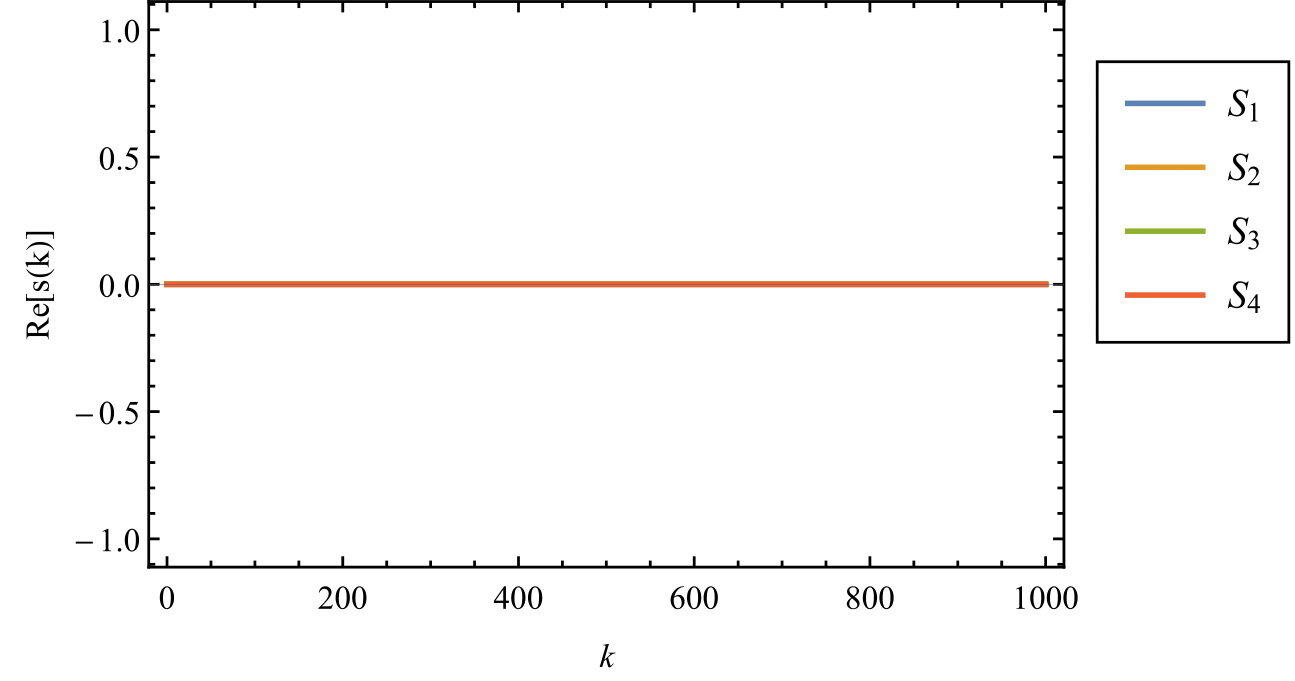}
\end{minipage}%
\begin{minipage}{.5\textwidth}
  \centering
  \includegraphics[height=4.6cm,width=1\linewidth]{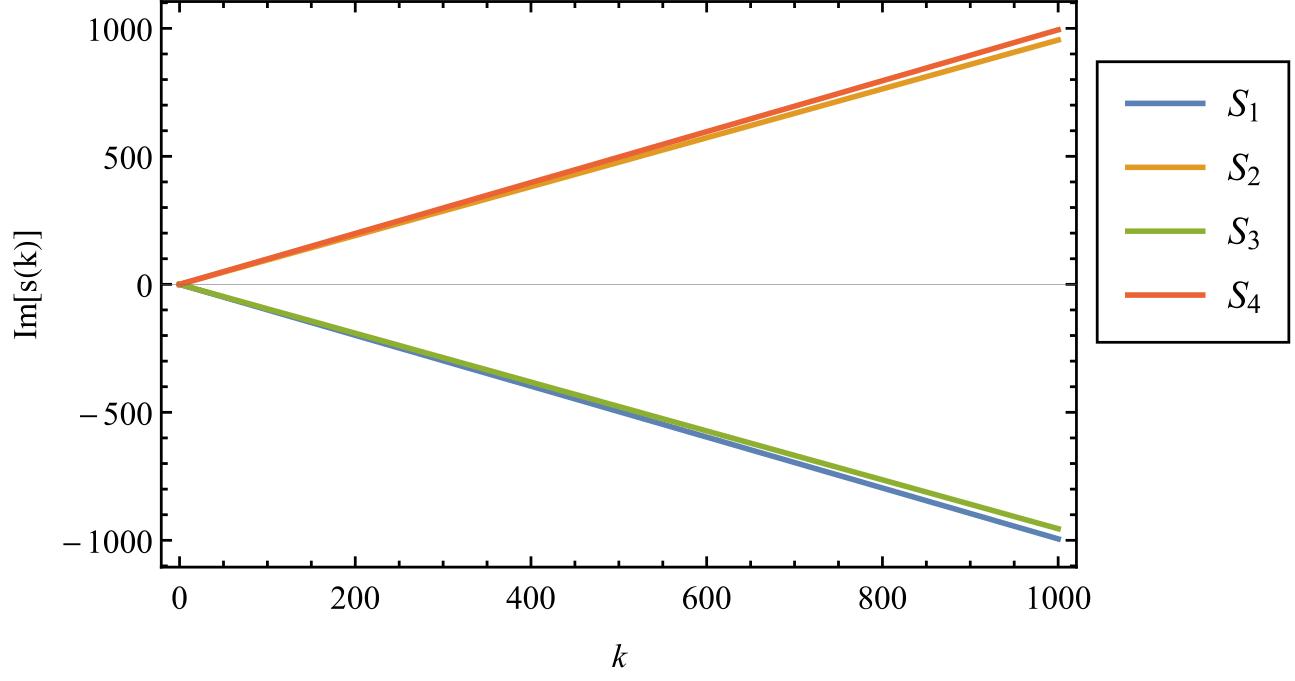}
\end{minipage}
\caption{\label{Fig:dCS_Toy1} (Color online) Real and imaginary parts of the dispersion relation solutions $s(k)$ for the model Eq.~(\ref{Eq:Model2}), with parameters $a=0.1$, $b=0.001$, $\lambda=0$. No blowing-up modes are present.}
\end{figure}

In the case of $\lambda=0$ and taking our usual phase function of the form $\phi(t,x)=st+ikx$, from Eq.~(\ref{Eq:Model2}) we find the relation
\begin{equation}
\left( s^2+k^2 \right)^2-ak^2 \left( s^2+k^2 \right)+bk^4=0,\label{toy1}
\end{equation}
which leads to the dispersion relation
\begin{equation}
s(k)^2=-k^2\Bigg{[}1-\frac{a}{2}\Bigg{(}1\pm\sqrt{1-\frac{4b}{a^2}}\Bigg{)} \Bigg{]}. \label{toy2}
\end{equation}
It is not difficult to choose pairs of values $(a,b)$ corresponding to nonblowing-up modes propagating at constant speeds. See for example Fig.~\ref{Fig:dCS_Toy1}.

\begin{figure}[h!]
\centering
\begin{minipage}{.5\textwidth}
  \centering
  \includegraphics[height=4.6cm,width=1\linewidth]{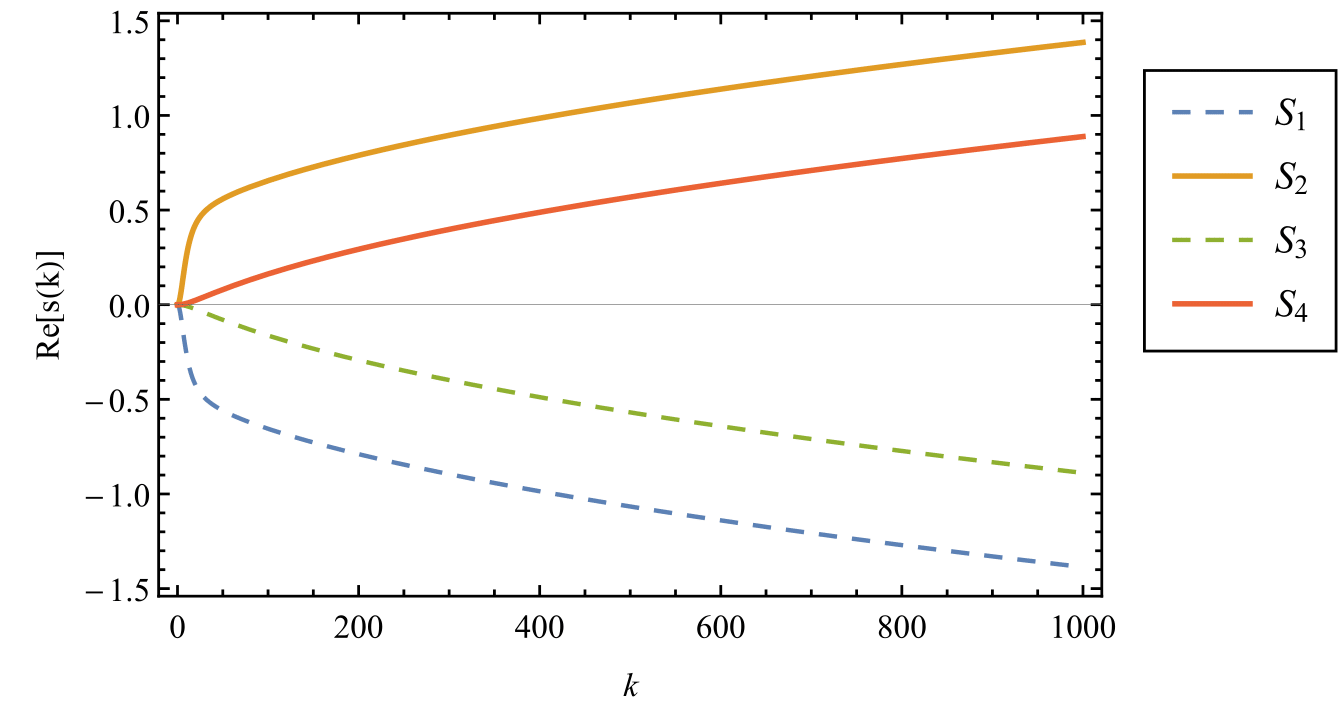}
\end{minipage}%
\begin{minipage}{.5\textwidth}
  \centering
  \includegraphics[height=4.6cm,width=1\linewidth]{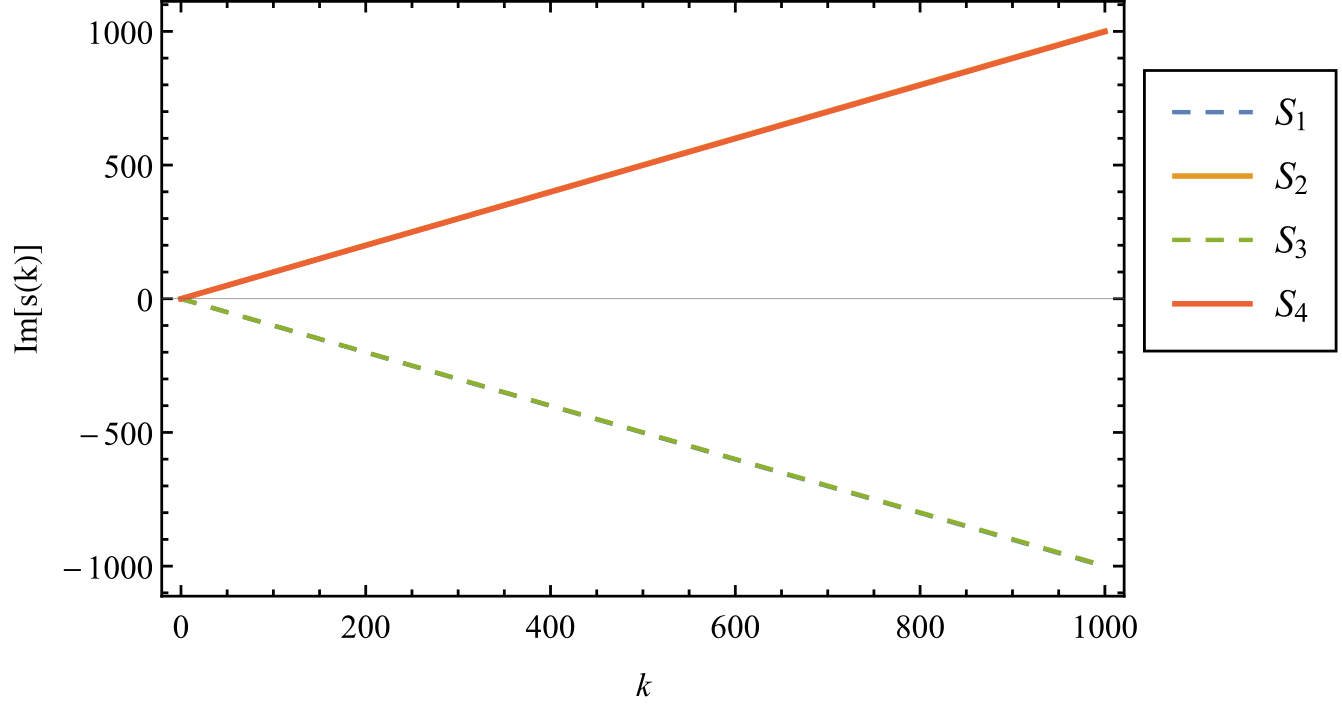}
\end{minipage}
\caption{\label{Fig:dCS_Toy2} (Color online) Real and imaginary parts of the dispersion relation solutions $s(k)$ for the model Eq.~(\ref{Eq:Model2}), with parameters $a=0.1$, $b=0.001$, $\lambda=0.1$. Positive, unbounded real parts reveal the existence of blowing-up modes.}
\end{figure}

We now turn on the high derivative term in Eq.~(\ref{Eq:Model2}) by taking $\lambda \ne 0$. The result is the appearance of positive, unbounded real parts for all of the solutions of the dispersion relation $s(k)$, indicating the presence of blowing-up modes for Eq.~(\ref{Eq:Model2}). See Fig.~\ref{Fig:dCS_Toy2}.

\begin{figure}[h!]
\centering
\begin{minipage}{.5\textwidth}
  \centering
  \includegraphics[height=4.8cm,width=1\linewidth]{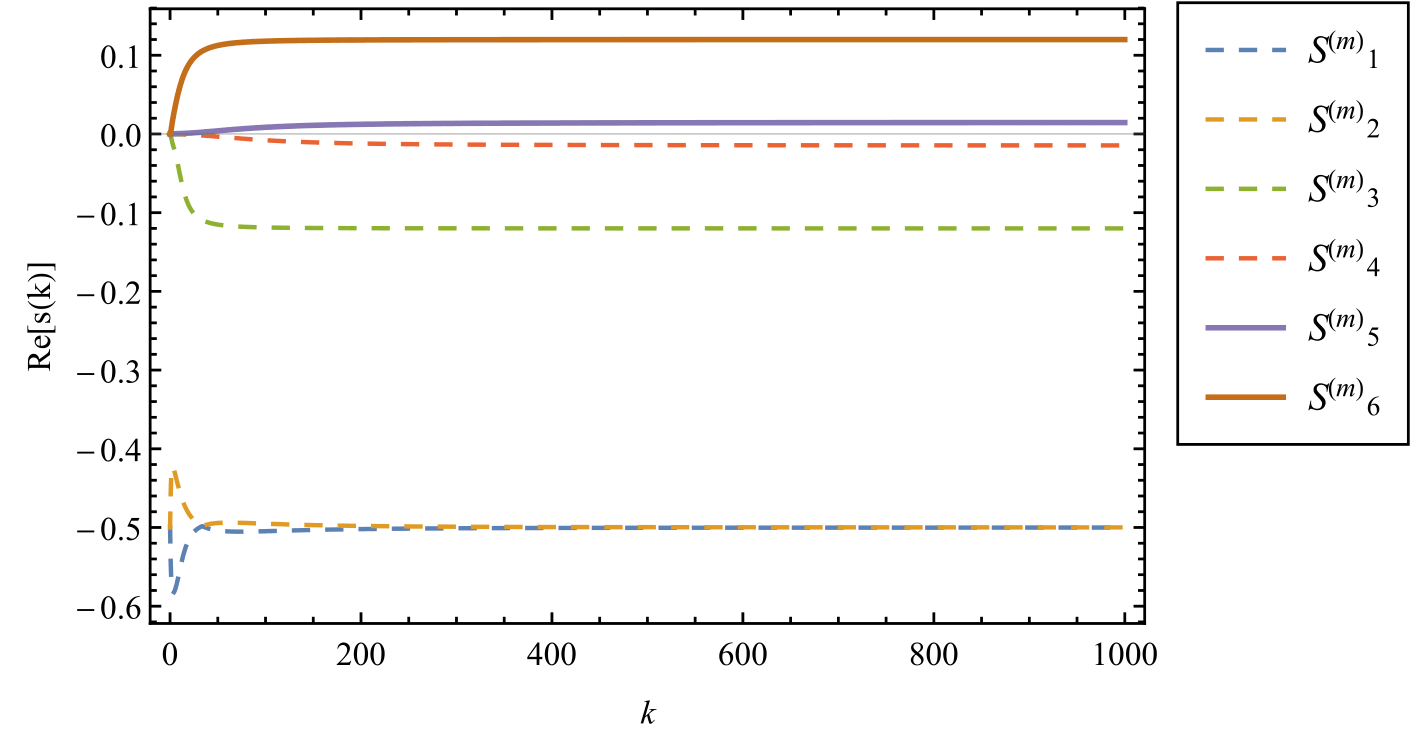}
\end{minipage}%
\begin{minipage}{.5\textwidth}
  \centering
  \includegraphics[height=4.8cm,width=1\linewidth]{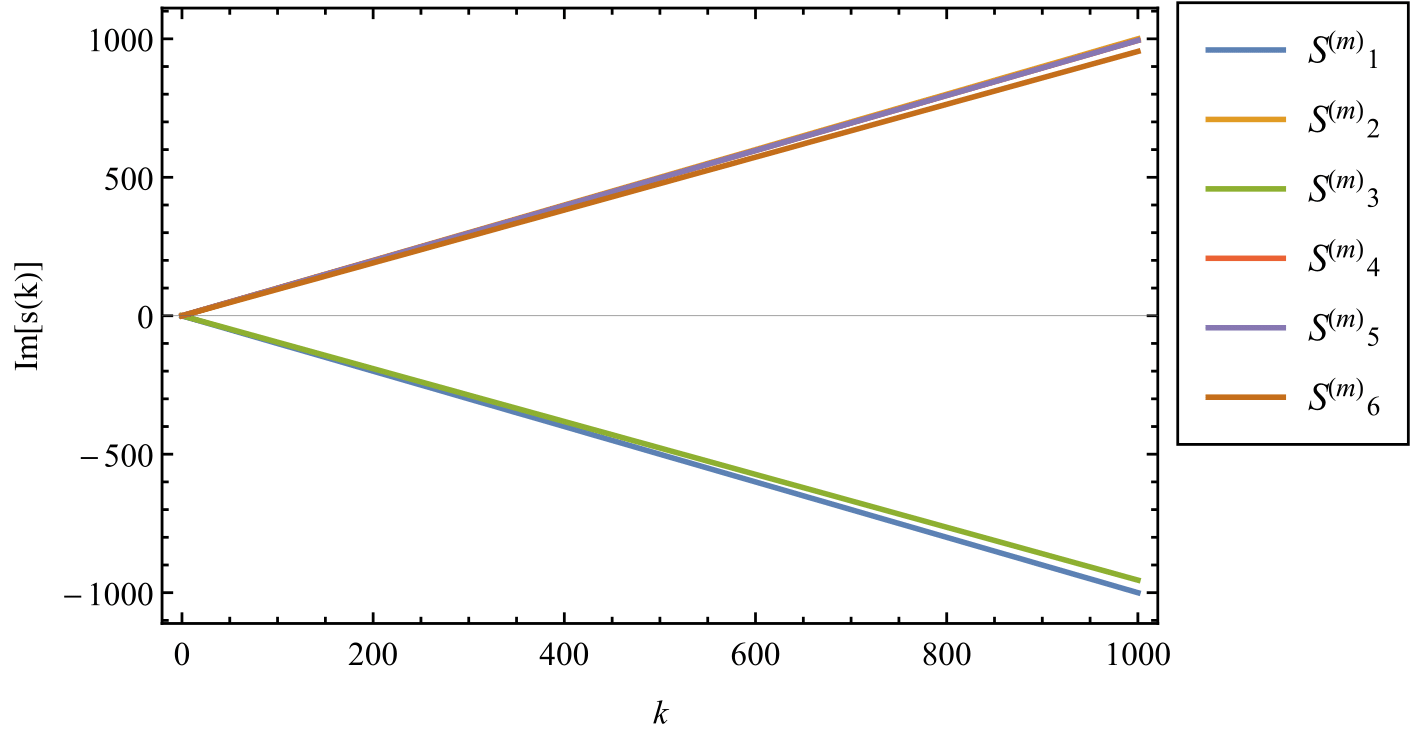}
\end{minipage}
\caption{\label{Fig:dCS_Toy3} (Color online) Real and imaginary parts of the dispersion relation solutions $s(k)$ for the system~(\ref{Eq:dCS_Toy_System}), with parameters $a=0.1$, $b=0.001$, $\lambda=0.1$, $\tau =0.5$ and $\sigma=0.5$. Real parts are bounded.}
\end{figure}

Following our program, we fix the problem revealed in Fig.~\ref{Fig:dCS_Toy2} by promoting the higher derivative in Eq.~(\ref{Eq:Model2}) to a new dynamical variable $\Pi:=\Box\Box\varphi$ obeying an evolution equation that will keep high order gradients small. The dynamics is now dictated by the system
\begin{subequations}\label{Eq:dCS_Toy_System}
\begin{eqnarray}
\Box\Box\varphi+a\partial_{x}^{2}\Box{\varphi}+b\partial^{4}_{x}\varphi+\lambda\partial_{x}\Pi &=& 0, \\
\tau\partial_{t}\Pi + \Pi - \Box\Box\varphi + \sigma\Box\Pi &=& 0.
\end{eqnarray}
\end{subequations}
Again with the ansatz $\varphi(t,x) = A\exp(s t + i k x)$, $\Pi(t,x) = A\beta(s,k)\exp(s t + i k x)$, system~(\ref{Eq:dCS_Toy_System}) implies the polynomial equation for $s(k)$
\begin{equation}\label{Eq:dCS_Toy_Relation}
\bigg{[}1 + \tau s(k) + \sigma\left(s(k)^2+k^2\right)\bigg{]}\bigg{[} \left(s(k)^2+k^2\right)^2 -ak^2\left(s(k)^2+k^2\right)+bk^4\bigg{]} +ik\lambda \left(s(k)^2+k^2\right)^2=0.
\end{equation}
Fig.~\ref{Fig:dCS_Toy3} provides evidence that, for parameters $\tau =0.5$, $\sigma=0.5$, and the same choice of $a=0.1$, $b=0.001$, $\lambda=0.1$, although the real part of some of the solutions of Eq.~(\ref{Eq:dCS_Toy_Relation}) can be non-negative, its growth is bounded in a significantly wide domain of wave numbers $k$. Such a result persists in a broad range of coefficients $(a,b,\lambda,\tau,\sigma)$.

Although exponentially growing modes are still present, the suggested existence of an upper bound for $\Re[s(k)]$ for all $k$, imply the existence of an upper bound in the solutions' rate of growth. In addition, mode propagation speeds are presumably subluminal, thus the problem of arbitrarily large mode propagation speeds of Eq.~(\ref{Eq:Model2}) is fixed as we suitably modulate the higher derivative term by virtue of the alternative system~(\ref{Eq:dCS_Toy_System}).

As the toy model Eq.~(\ref{Eq:Model2}) captures the structure of the linearized dCS field equations, our result suggests an strategy to alleviate the instability of high frequency modes in actual dCS scenarios.

\subsubsection{dCS field equations as a $3+1$ evolution system}\label{Sec:dCS3+1}

As pointed out by the authors of Ref.~\cite{Delsate_etal15}, besides the aspects of causality and finite propagation speeds of linearized dCS gravity, the question of well-posedness of the full set of dCS field equations written as a time evolution system is highly relevant. To the best of our knowledge, a rigorous mathematical proof of well-posedness of this system does not exist. Indeed, strong suggestions of ill-posedness in the most generic case can be inferred from the results of Ref.~\cite{Delsate_etal15}. 

A careful formulation of dCS field equations~(\ref{Eq:CS_field_eqns}) as a Cauchy problem can be found in Ref.~\cite{Delsate_etal15}, where a $3+1$ split of the equations is performed, auxiliary variables are defined, and a closed evolution system is written down for the spatial metric, the extrinsic curvature and the scalar field. The evolution system turns out to be rather cumbersome. However, the authors manage to construct a simple toy model capturing the structure of the actual system and the issues that suggest ill-posedness. Acausal mode propagation is present in this toy model, for instance.
Here we resume their toy model and explore its acausal pathology. We systematically implement our ideas to alleviate the issues.
The toy model by Delsate-Hilditch-Witek~\cite{Delsate_etal15} consists of a first order in time equation of the form\footnote{The model presented in Section IV-D-5 of Ref.~\cite{Delsate_etal15} can be cast into a constant coefficient system of the same structure as Eq.~(\ref{Eq:DHW_model}).}
\begin{equation}\label{Eq:DHW_model}
\dot{u}+u-u''=0.
\end{equation}
Considering again solutions of the form $u = A\exp(st+ikx)$, we find the dispersion relation
\begin{equation}
s(k) = -1-k^2.
\end{equation}
Even though $\Re[s(k)] < 0$ for arbitrary $k$, the phase velocity $s(k)/k$ grows arbitrarily large for modes in the high frequency regime, thus eventually violating causality.
In order to prevent this, let us define the new variable $\Pi := u'$ and build the alternative system
\begin{subequations}\label{Eq:CS_toy_system}
\begin{eqnarray}
\dot{u}&=&-u+\Pi',\\
\tau\dot{\Pi} &=& - \Pi + u',
\end{eqnarray}
\end{subequations}
with $0< \tau \leq 1$ so the second equation imposes $\Pi \rightarrow u'$ within a time scale given by $\tau$. The first-order system~(\ref{Eq:CS_toy_system}) is well posed if it is strongly hyperbolic~\cite{Gustafsson1995}, meaning that its principal symbol
\[
P := \left( \begin{array}{cc}
	0         & -1\\
	-1/\tau &  0
	\end{array}
\right),
\]
is diagonalizable with real eigenvalues. It is not difficult to check that the eigenvalues of $P$ are $\pm 1/\sqrt{\tau} \in \mathbb{R}$. Moreover, the set of eigenvectors $\left.\{ \left(1, -1/\sqrt{\tau}\right)^T, \left(1, 1/\sqrt{\tau}\right)^T \right.\}$ is linearly independent, then $P$ is diagonalizable. Therefore, the system~(\ref{Eq:CS_toy_system}) is well posed.
For completeness, one can verify that, writing $u = A\exp(st+ikx)$, $\Pi = A\beta(s,k)\exp(st+ikx)$, the system~(\ref{Eq:CS_toy_system}) has associated dispersion relation $s(k)$ satisfying
\begin{equation}
{\tau}s(k)^2+s(k)(\tau+1)+k^2+1 = 0,
\end{equation}
whose solutions are given by
\begin{equation}
s(k) = -\frac{\tau+1}{2\tau} \pm \frac{1}{2\tau}\sqrt{(\tau + 1)^2-4\tau(k^2+1)}.
\end{equation}
It is clear that $\Re[s(k)] \leq 0$ for all $k$. Furthermore, the phase velocity in this case is
\begin{equation}
\frac{s(k)}{k} = -\frac{\tau+1}{2\tau k} \pm \frac{1}{2\tau}\sqrt{\frac{(\tau - 1)^2}{k^2} - 4\tau}, \label{vel1}
\end{equation}
whose magnitude asymptotes $1/\sqrt{\tau} \leq 1$ as $k\rightarrow\infty$. As expected, this is consistent with the eigenvalues of the principal symbol $P$. Therefore, arbitrarily high frequency modes propagate at nonsuperluminal speeds.

\section{Final comments}\label{Sec:Conclusions}
In the current work we have presented a rather general approach to address pathologies typically found in many extensions to General Relativity and their applications in the strongly nonlinear/highly dynamical systems described by compact binary mergers.  Our approach relies on controlling the behavior of energy cascade to high frequencies so that the norm of the solution can be bounded and problems studied with them are rendered well posed. Additionally, we have argued that such an approach applied to 3+1 dimensional systems would yield solutions which are faithful to the principles which underlie the original theory. That such a behavior happens in the long-wavelength regime is still not solidly established, but is supported by several tantalizing observations, including the recent detections of gravitational waves. This opens the door to implementing this approach in particular theories of interest and study their consequences for gravitational wave astronomy. As more detections are available---together with future, more sensitive, facilities---important constraints on possibly subtle deviations can be potentially obtained.

Is our approach the only viable option? Unlikely, as several approaches can be envisioned. These include, for instance, an ``iterative'' strategy where the zeroth order solution is provided by General Relativity and this solution is then employed to evaluate the ``corrections''~\cite{Endlich:2017tqa}. This strategy would require saving the solution to then redo the problem with new sources and iterate as needed. Alternatively, a related strategy could be turned into a hierarchical strategy which effectively enlarges the fields to solve for but does not require saving a particular step in the approach  (e.g.~\cite{Okounkova:2017yby}). The strategies mentioned above would certainly address the ill-posedness concern but may require several iterations (or members of the perturbative hierarchy computed) to capture the correct physical behavior. It is possible that the cost associated in implementing these options is high but, crucially, it might lead to secular modes which might require a suitable resummation to be implemented. Notwithstanding these possible issues, there is {\it a priori} no guarantee that the correcting terms will remain small {\em unless} one can argue there is no significant energy cascade towards the ultraviolet regime. Fortunately for this question, the very same observation pointed here---as to the possibly mainly inverse energy cascade in 3+1 dimensions for suitable regimes of interest---would help to ease, to some degree, this concern. We note that other efforts have focused on healing pathologies in specific theories, in
particular the removal of ghost degrees of freedom~\cite{tCmFeLaT13}. However, removal of ghosts is no guarantee that a theory would admit a well posed formulation (e.g.~\cite{Papallo:2017qvl}). Of course, the existence of ghosts implies ill-posedness, but not the inverse.

\acknowledgments
We thank Enrico Barausse, Cliff Burgess, William East, Stephen Green, Anna Ijjas, Djordje Minic, Rob Myers, Frans Pretorius, Jorge Pullin, Oscar Reula, Thomas Sotiriou, Leo Stein, Mark Trodden, Nicol\'as Yunes, David Garfinkle, Eric Poisson, Andrew Tolley, and Olivier Sarbach for interesting discussions. Research at the Perimeter Institute is supported by the Government of Canada through Industry Canada and by the Province of Ontario through the Ministry of Research and Innovation.

\bibliographystyle{unsrt}
\bibliography{references}

\begin{thebibliography}{10}

\bibitem{Will:2014bqa}
Clifford~M. Will.
\newblock {Was Einstein Right? A Centenary Assessment}.
\newblock 2014.
\newblock arXiv:1409.7871.

\bibitem{Freire:2012mg}
Paulo C.~C. Freire, Norbert Wex, Gilles Esposito-Farese, Joris P.~W. Verbiest,
  Matthew Bailes, Bryan~A. Jacoby, Michael Kramer, Ingrid~H. Stairs, John
  Antoniadis, and Gemma~H. Janssen.
\newblock {The relativistic pulsar-white dwarf binary PSR J1738+0333 II. The
  most stringent test of scalar-tensor gravity}.
\newblock {\em Mon. Not. Roy. Astron. Soc.}, 423:3328, 2012.

\bibitem{Baker:2012zs}
Tessa Baker, Pedro~G. Ferreira, and Constantinos Skordis.
\newblock {The Parameterized Post-Friedmann framework for theories of modified
  gravity: concepts, formalism and examples}.
\newblock {\em Phys. Rev.}, D87(2):024015, 2013.

\bibitem{Yunes_Siemens13}
Nicol{\'a}s Yunes and Xavier Siemens.
\newblock Gravitational-wave tests of general relativity with ground-based
  detectors and pulsar-timing arrays.
\newblock {\em Living Reviews in Relativity}, 16(1):9, 2013.

\bibitem{Berti_etal15}
Emanuele Berti, Enrico Barausse, Vitor Cardoso, Leonardo Gualtieri, Paolo Pani,
  Ulrich Sperhake, Leo~C Stein, Norbert Wex, Kent Yagi, Tessa Baker, C~P
  Burgess, Fl\'avio~S Coelho, Daniela Doneva, Antonio~De Felice, Pedro~G
  Ferreira, Paulo C~C Freire, James Healy, Carlos Herdeiro, Michael Horbatsch,
  Burkhard Kleihaus, Antoine Klein, Kostas Kokkotas, Jutta Kunz, Pablo Laguna,
  Ryan~N Lang, Tjonnie G~F Li, Tyson Littenberg, Andrew Matas, Saeed
  Mirshekari, Hirotada Okawa, Eugen Radu, Richard O'Shaughnessy, Bangalore~S
  Sathyaprakash, Chris Van~Den Broeck, Hans~A Winther, Helvi Witek, Mir~Emad
  Aghili, Justin Alsing, Brett Bolen, Luca Bombelli, Sarah Caudill, Liang Chen,
  Juan~Carlos Degollado, Ryuichi Fujita, Caixia Gao, Davide Gerosa, Saeed
  Kamali, Hector~O Silva, Joao~G Rosa, Laleh Sadeghian, Marco Sampaio, Hajime
  Sotani, and Miguel Zilhao.
\newblock Testing general relativity with present and future astrophysical
  observations.
\newblock {\em Classical and Quantum Gravity}, 32(24):243001, 2015.

\bibitem{Abbott_etal16}
B.~P.~Abbott {\it et al}.~(LIGO Scientific~Collaboration and Virgo
  Collaboration).
\newblock Binary black hole mergers in the first advanced ligo observing run.
\newblock {\em Phys. Rev. X}, 6:041015, Oct 2016.

\bibitem{Yunes_etal16}
Nicol\'as Yunes, Kent Yagi, and Frans Pretorius.
\newblock Theoretical physics implications of the binary black-hole mergers
  gw150914 and gw151226.
\newblock {\em Phys. Rev. D}, 94:084002, Oct 2016.

\bibitem{Cornish:2011ys}
Neil Cornish, Laura Sampson, Nicol\'as Yunes, and Frans Pretorius.
\newblock {Gravitational Wave Tests of General Relativity with the
  Parameterized Post-Einsteinian Framework}.
\newblock {\em Phys. Rev.}, D84:062003, 2011.

\bibitem{Meidam:2014jpa}
J.~Meidam, M.~Agathos, C.~Van Den~Broeck, J.~Veitch, and B.~S. Sathyaprakash.
\newblock {Testing the no-hair theorem with black hole ringdowns using TIGER}.
\newblock {\em Phys. Rev.}, D90(6):064009, 2014.

\bibitem{Sampson:2014qqa}
Laura Sampson, Nicol\'as Yunes, Neil Cornish, Marcelo Ponce, Enrico Barausse,
  Antoine Klein, Carlos Palenzuela, and Luis Lehner.
\newblock {Projected Constraints on Scalarization with Gravitational Waves from
  Neutron Star Binaries}.
\newblock {\em Phys. Rev.}, D90(12):124091, 2014.

\bibitem{Yang:2017zxs}
Huan Yang, Kent Yagi, Jonathan Blackman, Luis Lehner, Vasileios Paschalidis,
  Frans Pretorius, and Nicol\'as Yunes.
\newblock Black hole spectroscopy with coherent mode stacking.
\newblock {\em Phys. Rev. Lett.}, 118:161101, Apr 2017.

\bibitem{jH02}
Jacques Hadamard.
\newblock {Sur les probl\`emes aux d\'eriv\'ees partielles et leur
  signification physique}.
\newblock {\em Princeton University Bulletin}, 13:49--52, 1902.

\bibitem{Gustafsson1995}
Bertil Gustafsson, Heinz-Otto Kreiss, and Joseph Oliger.
\newblock {\em Time dependent problems and difference methods}.
\newblock John Wiley \& Sons Inc., 1995.

\bibitem{Papallo:2017qvl}
Giuseppe Papallo and Harvey~S. Reall.
\newblock On the local well-posedness of lovelock and horndeski theories.
\newblock {\em Phys. Rev. D}, 96:044019, Aug 2017.

\bibitem{Endlich:2017tqa}
Solomon Endlich, Victor Gorbenko, Junwu Huang, and Leonardo Senatore.
\newblock An effective formalism for testing extensions to general relativity
  with gravitational waves.
\newblock {\em Journal of High Energy Physics}, 2017(9):122, Sep 2017.

\bibitem{Okounkova:2017yby}
Maria Okounkova, Leo~C. Stein, Mark~A. Scheel, and Daniel~A. Hemberger.
\newblock Numerical binary black hole mergers in dynamical chern-simons
  gravity: Scalar field.
\newblock {\em Phys. Rev. D}, 96:044020, Aug 2017.

\bibitem{Sayantani_etal08}
Sayantani Bhattacharyya, Shiraz Minwalla, Veronika~E. Hubeny, and Mukund
  Rangamani.
\newblock Nonlinear fluid dynamics from gravity.
\newblock {\em Journal of High Energy Physics}, 2008(02):045, 2008.

\bibitem{Sayantani_etal08b}
Sayantani Bhattacharyya, R.~Loganayagam, Ipsita Mandal, Shiraz Minwalla, and
  Ankit Sharma.
\newblock Conformal nonlinear fluid dynamics from gravity in arbitrary
  dimensions.
\newblock {\em Journal of High Energy Physics}, 2008(12):116, 2008.

\bibitem{VanRaamsdonk08}
Mark~Van Raamsdonk.
\newblock Black hole dynamics from atmospheric science?
\newblock {\em Journal of High Energy Physics}, 2008(05):106, 2008.

\bibitem{Yang_etal15}
Huan Yang, Aaron Zimmerman, and Luis Lehner.
\newblock Turbulent black holes.
\newblock {\em Phys. Rev. Lett.}, 114:081101, Feb 2015.

\bibitem{Yang:2015jja}
Huan Yang, Fan Zhang, Stephen~R. Green, and Luis Lehner.
\newblock {Coupled Oscillator Model for Nonlinear Gravitational Perturbations}.
\newblock {\em Phys. Rev.}, D91(8):084007, 2015.

\bibitem{Galtier:2017mve}
S\'ebastien Galtier and Sergey~V. Nazarenko.
\newblock {Turbulence of Weak Gravitational Waves in the Early Universe}.
\newblock 2017.
\newblock arXiv:1703.09069.

\bibitem{Baier_etal08}
Rudolf Baier, Paul Romatschke, Dam~Thanh Son, Andrei~O. Starinets, and
  Mikhail~A. Stephanov.
\newblock Relativistic viscous hydrodynamics, conformal invariance, and
  holography.
\newblock {\em Journal of High Energy Physics}, 2008(04):100, 2008.

\bibitem{Romatschke10}
Paul Romatschke.
\newblock New developments in relativistic viscous hydrodynamics.
\newblock {\em International Journal of Modern Physics E}, 19(01):1--53, 2010.

\bibitem{Buchel:2009tt}
Alex Buchel and Robert~C. Myers.
\newblock {Causality of Holographic Hydrodynamics}.
\newblock {\em JHEP}, 08:016, 2009.

\bibitem{Israel76}
Werner Israel.
\newblock Nonstationary irreversible thermodynamics: A causal relativistic
  theory.
\newblock {\em Annals of Physics}, 100(1–2):310 -- 331, 1976.

\bibitem{Israel_Stewart76}
W.~Israel and J.M. Stewart.
\newblock Thermodynamics of nonstationary and transient effects in a
  relativistic gas.
\newblock {\em Physics Letters A}, 58(4):213 -- 215, 1976.

\bibitem{Israel_Stewart79}
W.~Israel and J.M. Stewart.
\newblock Transient relativistic thermodynamics and kinetic theory.
\newblock {\em Annals of Physics}, 118(2):341 -- 372, 1979.

\bibitem{Jou_etal98}
D~Jou, J~Casas-Vazquez, and G~Lebon.
\newblock Extended irreversible thermodynamics.
\newblock {\em Reports on Progress in Physics}, 51(8):1105, 1988.

\bibitem{Kraichnan67}
Robert~H. Kraichnan.
\newblock Inertial ranges in two-dimensional turbulence.
\newblock {\em The Physics of Fluids}, 10(7):1417--1423, 1967.

\bibitem{Carrasco_etal12}
Federico Carrasco, Luis Lehner, Robert~C. Myers, Oscar Reula, and Ajay Singh.
\newblock Turbulent flows for relativistic conformal fluids in $2\mathbf{+}1$
  dimensions.
\newblock {\em Phys. Rev. D}, 86:126006, Dec 2012.

\bibitem{Westernacher-Schneider_Lehner15}
John~Ryan Westernacher-Schneider, Luis Lehner, and Yaron Oz.
\newblock Scaling relations in two-dimensional relativistic hydrodynamic
  turbulence.
\newblock {\em Journal of High Energy Physics}, 2015(12):67, 2015.

\bibitem{Green_etal14}
Stephen~R. Green, Federico Carrasco, and Luis Lehner.
\newblock Holographic path to the turbulent side of gravity.
\newblock {\em Phys. Rev. X}, 4:011001, Jan 2014.

\bibitem{Hiscock_and_Lindblom83}
William~A Hiscock and Lee Lindblom.
\newblock Stability and causality in dissipative relativistic fluids.
\newblock {\em Annals of Physics}, 151(2):466 -- 496, 1983.

\bibitem{doi:10.1063/1.530958}
Robert Geroch.
\newblock Relativistic theories of dissipative fluids.
\newblock {\em Journal of Mathematical Physics}, 36(8):4226--4241, 1995.

\bibitem{Eling_etal10}
Christopher Eling, Itzhak Fouxon, and Yaron Oz.
\newblock {Gravity and a Geometrization of Turbulence: An Intriguing
  Correspondence}.
\newblock 2010.
\newblock arXiv:1004.2632.

\bibitem{Adams_etal14}
Allan Adams, Paul~M. Chesler, and Hong Liu.
\newblock Holographic turbulence.
\newblock {\em Phys. Rev. Lett.}, 112:151602, Apr 2014.

\bibitem{Westernacher2}
John~Ryan Westernacher-Schneider and Luis Lehner.
\newblock Numerical measurements of scaling relations in two-dimensional
  conformal fluid turbulence.
\newblock {\em Journal of High Energy Physics}, 2017(8):27, Aug 2017.

\bibitem{Abbott:2016blz}
B.~ P. Abbott et~al.
\newblock {Observation of Gravitational Waves from a Binary Black Hole Merger}.
\newblock {\em Phys. Rev. Lett.}, 116(6):061102, 2016.

\bibitem{Abbott:2016nmj}
B. P. Abbott et~al.
\newblock {GW151226: Observation of Gravitational Waves from a 22-Solar-Mass
  Binary Black Hole Coalescence}.
\newblock {\em Phys. Rev. Lett.}, 116(24):241103, 2016.

\bibitem{Abbott:2017}
B.~ P. Abbott et~al.
\newblock Gw170104: Observation of a 50-solar-mass binary black hole
  coalescence at redshift 0.2.
\newblock {\em Phys. Rev. Lett.}, 118:221101, Jun 2017.

\bibitem{Christodoulou_Klainerman93}
Demetrios {Christodoulou} and Sergiu {Klainerman}.
\newblock {\em {The global nonlinear stability of the Minkowski space.}}
\newblock Princeton, NJ: Princeton University Press, 1993.

\bibitem{Lindblad_Rodnianski10}
Hans {Lindblad} and Igor {Rodnianski}.
\newblock {The global stability of Minkowski space-time in harmonic gauge.}
\newblock {\em {Ann. Math. (2)}}, 171(3):1401--1477, 2010.

\bibitem{1965PhRvL..14...57P}
R.~{Penrose}.
\newblock {Gravitational Collapse and Space-Time Singularities}.
\newblock {\em Physical Review Letters}, 14:57--59, January 1965.

\bibitem{Hawking:1973uf}
S.~W. Hawking and G.~F.~R. Ellis.
\newblock {\em {The Large Scale Structure of Space-Time}}.
\newblock Cambridge Monographs on Mathematical Physics. Cambridge University
  Press, 2011.

\bibitem{mC93}
M.W. Choptuik.
\newblock Universality and scaling in gravitational collapse of a massless
  scalar field.
\newblock {\em Phys.Rev.Lett.}, 70:9--12, 1993.

\bibitem{NCG:94}
Alain Connes.
\newblock {\em Noncommutative Geometry}.
\newblock Academic Press, New York, 1994.

\bibitem{NCG2:08}
A.~Connes and M.~Marcolli.
\newblock {\em Noncommutative Geometry, Quantum Fields and Motives}.
\newblock Hindustan Book Agency, India, 2008.

\bibitem{Sakellariadou:2011dk}
Mairi Sakellariadou.
\newblock {Cosmology within Noncommutative Spectral Geometry}.
\newblock {\em PoS}, CNCFG2010:028, 2010.

\bibitem{Jackiw_and_Pi03}
R.~Jackiw and S.-Y. Pi.
\newblock {C}hern-{S}imons modification of general relativity.
\newblock {\em Phys. Rev. D}, 68:104012, Nov 2003.

\bibitem{Alexander_Yunes09}
Stephon Alexander and Nicol\'as Yunes.
\newblock {C}hern-{S}imons modified general relativity.
\newblock {\em Physics Reports}, 480(1–2):1 -- 55, 2009.

\bibitem{Polchinski98}
J.~Polchinski.
\newblock {\em String Theory: Volume 2, Superstring Theory and Beyond}.
\newblock Cambridge Monographs on Mathematical Physics. Cambridge University
  Press, 1998.

\bibitem{Grumiller_etal08}
Daniel Grumiller, Robert Mann, and Robert McNees.
\newblock Dirichlet boundary-value problem for {C}hern-{S}imons modified
  gravity.
\newblock {\em Phys. Rev. D}, 78:081502, Oct 2008.

\bibitem{Delsate_etal15}
T\'erence Delsate, David Hilditch, and Helvi Witek.
\newblock Initial value formulation of dynamical {C}hern-{S}imons gravity.
\newblock {\em Phys. Rev. D}, 91:024027, Jan 2015.

\bibitem{dGfPnY10}
David Garfinkle, Frans Pretorius, and Nicol\'as Yunes.
\newblock Linear stability analysis and the speed of gravitational waves in
  dynamical {C}hern-{S}imons modified gravity.
\newblock {\em Phys. Rev. D}, 82:041501, Aug 2010.

\bibitem{aDkYnY14}
Dimitry Ayzenberg, Kent Yagi, and Nicol\'as Yunes.
\newblock Linear stability analysis of dynamical quadratic gravity.
\newblock {\em Phys. Rev. D}, 89:044023, Feb 2014.

\bibitem{hMtS14}
Hayato Motohashi and Teruaki Suyama.
\newblock Black hole perturbation in nondynamical and dynamical
  {C}hern-{S}imons gravity.
\newblock {\em Phys. Rev. D}, 85:044054, Feb 2012.

\bibitem{dBjmMGgaMM09}
David Brizuela, Jos{\'e}~M. Mart{\'i}n-Garc{\'i}a, and Guillermo~A.
  Mena~Marug{\'a}n.
\newblock x{P}ert: computer algebra for metric perturbation theory.
\newblock {\em General Relativity and Gravitation}, 41(10):2415, 2009.

\bibitem{xPert}
x{P}ert.
\newblock \url{http://www.xact.es/}.

\bibitem{mathematica}
{W}olfram~{R}esearch {I}nc.
\newblock {M}athematica 11.1.1.0.
\newblock {\em {C}hampaign, Illinois}, 2017.

\bibitem{tCmFeLaT13}
Tai-jun Chen, Matteo Fasiello, Eugene~A. Lim, and Andrew~J. Tolley.
\newblock {Higher derivative theories with constraints: Exorcising
  Ostrogradski's Ghost}.
\newblock {\em JCAP}, 1302:042, 2013.

\end{thebibliography}

\end{document}